\documentclass{aa}  
\usepackage{graphicx}
\usepackage{txfonts}
\usepackage{hyperref}
\usepackage{soul}

\usepackage{xcolor}
\usepackage[normalem]{ulem}
\usepackage{orcidlink}

\usepackage{subfigure}

\begin{document} 

\title{JWST observations of 47 Tucanae}
\subtitle{I. The low-mass main sequence down to the hydrogen-burning limit}

\author{M.\,Scalco\inst{1}\fnmsep\thanks{\email{mscalco@iu.edu}}\orcidlink{0000-0001-8834-3734}
\and 
R.\,Gerasimov\inst{2}\orcidlink{0000-0003-0398-639X}
\and
L.\,R.\,Bedin\inst{3}\orcidlink{0000-0003-4080-6466}
\and
E.\,Vesperini\inst{1}\orcidlink{0000-0003-2742-6872}
\and
M.\,Griggio\inst{4}\orcidlink{0000-0002-5060-1379}
\and
D.\,Nardiello\inst{5,3}\orcidlink{0000-0003-1149-3659}
\and
J.\,Anderson\inst{4}
}

\institute{
Department of Astronomy, Indiana University, Swain West, 727 E. 3rd Street, Bloomington, IN 47405, USA
\and
Department of Physics and Astronomy, University of Notre Dame, Nieuwland Science Hall, Notre Dame, IN 46556, USA
\and
Istituto Nazionale di Astrofisica, Osservatorio Astronomico di Padova, Vicolo dell'Osservatorio 5, Padova 35122, Italy
\and
Space Telescope Science Institute, 3700 San Martin Drive, Baltimore, MD 21218, USA
\and
Dipartimento di Fisica e Astronomia ``Galileo Galilei'', Universit{\`a} di Padova, Vicolo dell'Osservatorio 3, Padova 35122, Italy
}

\date{Received 12 April 2026 / Accepted 22 June 2026}
 
\abstract
{We present the first results of a series of studies based on \textit{James Webb} Space Telescope (JWST) observations of the globular cluster (GC) 47\,Tucanae (47\,Tuc, or NGC\,104) obtained with the Near Infrared Camera (NIRCam). By combining new data with earlier JWST observations, we derive high-precision proper motions (PMs) over a multi-year baseline (of $\sim3$\,years) and construct a clean sample of cluster members extending to very faint magnitudes ($m_{\rm F444W}\sim24.25$\,mag, corresponding to $\sim0.077\,M_{\odot}$, slightly above the estimated hydrogen-burning limit of 47\,Tuc, $\sim0.074\,M_{\odot}$) along the main sequence (MS). We investigated the presence of the low-mass MS kink previously identified in 47\,Tuc using the new F090W and F444W photometry. No significant signature of the feature is found using these filters, consistent with its proposed physical interpretation as a consequence of the transition in carbon chemistry toward a methane-dominated equilibrium in very low-mass stars. A comparison with theoretical isochrones indicates that the stellar population near the bottom of the MS is likely dominated by oxygen-poor stars, in agreement with previous findings. Using artificial-star tests, we quantify the completeness of our sample and show that the requirement of PM-based membership significantly limits the detection of the faintest objects, indicating that current JWST observations of 47\,Tuc are not sufficiently deep to reliably trace the cluster’s brown dwarf (BD) cooling sequence. The mass function (MF) of 47\,Tuc shows a structured behaviour, with a break at $\sim$0.15\,$M_{\odot}$ not associated with the onset of convection, and a mild but systematic radial variation consistent with mass segregation. Our analysis indicates that the MF shape in this mass regime is more complex than a simple power-law description. This feature in the MF may point to a presently unrecognised mechanism affecting the formation or dynamical evolution of low-mass stars in globular clusters.} 


\titlerunning{47\,Tuc low-mass main sequence with JWST}
\authorrunning{M.\,Scalco et al.}
\maketitle

\section{Introduction}\label{Section1}
Globular clusters (GCs) have long been recognised as fundamental laboratories for studying stellar evolution and dynamical processes in dense stellar systems. While the most massive stars in GCs have been extensively investigated over the past decades, the lowest-mass stars remain comparatively poorly explored. This is primarily due to their intrinsic faintness and to their typical distances of a few kiloparsecs \citep{1996AJ....112.1487H,2010arXiv1012.3224H,2021MNRAS.505.5957B}, which make the detection of very faint objects extremely challenging even with the most advanced facilities, including the \textit{Hubble} Space Telescope \citep[HST; see e.g.][]{2005AJ....130..626K,2008AJ....135.2141R,2012AJ....143...11K,2016ApJ...817...48D,2019MNRAS.486.2254D}.

Despite these observational challenges, the low-mass end of the main sequence (MS) offers unique insight into stellar physics and cluster evolution. The photometric colours of very low-mass stars are highly sensitive to variations in elemental abundances, owing to the dominant role of molecular chemistry and opacity in their cool atmospheres \citep{1986ApJ...310..238L,1996A&A...305L...1T,1995ApJ...445..433A,2002ApJ...568..335M,2016ApJ...823..102C,2020A&A...637A..38P,2022csss.confE.120G,2024ApJ...961..139G}.

At the same time, stars below \(\sim0.35\,M_\odot\) are expected to be fully convective and experience minimal internal chemical processing, as a result of the long timescales associated with hydrogen burning \citep{1997A&A...327.1039C,2018MNRAS.480.1711M,2018A&A...619A.177B,2021ApJ...907...53F,2021A&A...650A.184M,2022ApJ...930...24G,2023ApJ...944..129B}. Their surface abundances largely preserve the chemical composition of the gas from which they formed, making them valuable tracers of the formation and chemical evolution of GCs.

At fainter magnitudes, the stellar sequence transitions into the substellar regime populated by brown dwarfs (BDs), which have masses below the minimum required to sustain stable hydrogen fusion \citep[the hydrogen-burning limit, HBL, $\gtrsim$0.07$\,M_\odot$ for solar composition][]{1962AJ.....67S.579K,1963PThPh..30..460H,1963ApJ...137.1121K,1994ApJ...424..333S}. This transition gives rise to a characteristic feature in the luminosity function (LF), commonly referred to as the stellar–substellar gap \citep{1997A&A...328...83C,1998ApJ...509..192S,2004ApJS..155..191B,2024ApJ...961..139G}. The properties of this gap, including its depth and extent, are closely related to the age of the cluster, making the LF a potentially powerful age diagnostic.

\begin{centering} 
\begin{figure}
 \includegraphics[width=\columnwidth]{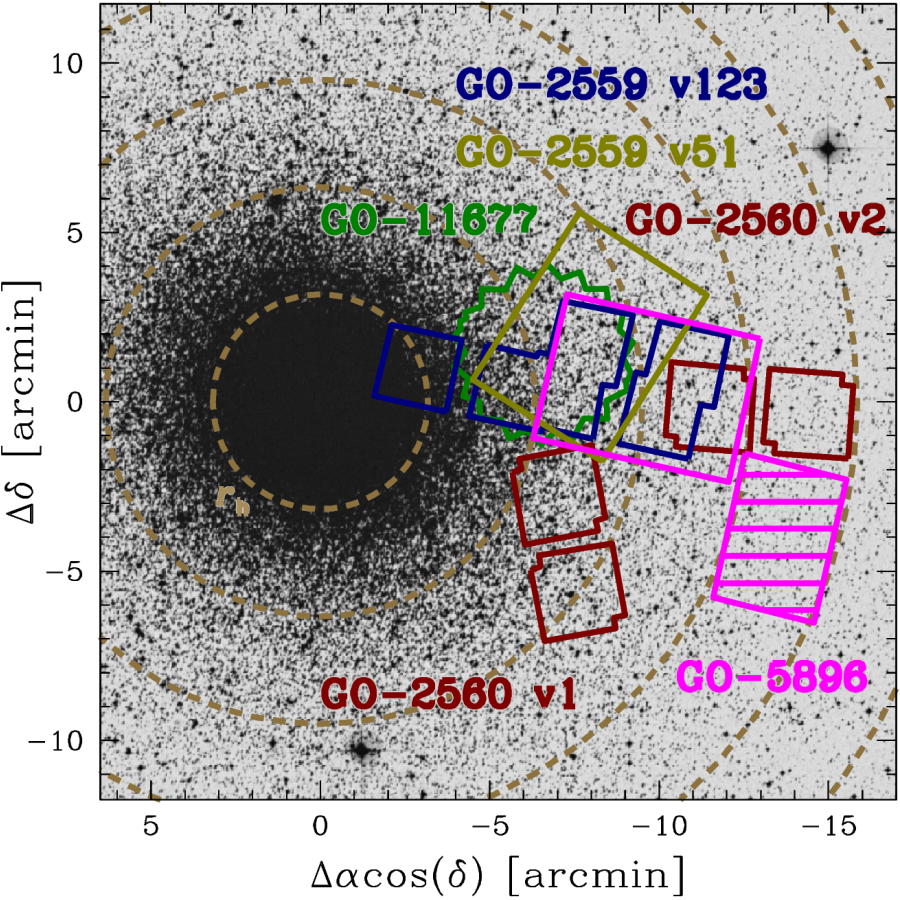}
 \caption{Footprints of the JWST GO-5896 observations (magenta) overlaid on a Digital Sky Survey image of 47\,Tuc. The NIRCam field of view is shown as an outlined footprint, while the corresponding NIRISS parallel field is indicated by the hatched footprint. For comparison, we also show the NIRCam fields observed under the GO-2559 programme (first epoch GO-2559 v123 in blue and second epoch GO-2559 v51 in yellow), the NIRCam pointings from the GO-2560 programme (visits v1 and v2 in red), and the Advanced Camera for Surveys/Wide Field Channel (ACS/WFC) field from the HST GO-11677 programme (in green). Axes are given in arcminutes from the cluster centre. The orange dashed circles mark the half-light radius \citep[$r_{\rm h}=3\farcm17$;][]{1996AJ....112.1487H,2010arXiv1012.3224H} and its multiples, $n\,r_{\rm h}$ with $n=2,\ldots,6$.} 
 \label{FOV} 
\end{figure} 
\end{centering}

The present-day GC mass function (MF) provides key constraints on both the initial conditions of star formation and the subsequent effects of dynamical evolution \citep{1993AJ....105.2148D,1997MNRAS.289..898V,2003MNRAS.340..227B,2017MNRAS.471.3668S,2017MNRAS.471.3845W,2018MNRAS.476.2731V,2020MNRAS.494.4226E,2020MNRAS.499.2390C,2020MNRAS.491..113H,2023MNRAS.522.5320D}. Extending MF measurements down to the very low-mass regime, and into the substellar domain, is therefore essential for obtaining a complete census of the cluster mass spectrum. Furthermore, in GCs hosting multiple stellar populations, differences in the MF among populations at low stellar masses may arise from variations in their initial mass distributions or from their distinct dynamical evolution, offering an additional probe of the formation and assembly history of these systems \citep[see][and references therein]{2024AN....34540018S,2024A&A...690A.371L}.

While thousands of BDs of solar metallicity have been identified and studied in the solar neighbourhood and in nearby young clusters and associations \citep[see e.g.][]{1995Natur.378..463N,1995Natur.377..129R,1996ApJ...458..600B,2010A&A...510A..27B,2012ApJ...753..156K,2019ApJS..240...19K,2021ApJ...915L...6K,2021AJ....161...42B}, the substellar regime of GCs remains much less explored. The advent of the \textit{James Webb} Space Telescope (JWST) has opened a new observational window on the low-mass regime of GCs. Its unprecedented sensitivity and angular resolution in the near- and mid-infrared allow the MS to be traced to significantly lower masses than previously accessible, approaching the HBL and the BD regime. At the same time, JWST provides high-precision astrometry. As JWST has now been operating for several years and has accumulated repeated observations of the same fields, the available temporal baselines are becoming sufficiently long to allow precise proper-motion (PM) studies based entirely on JWST data, enabling robust membership determination even at faint magnitudes.

The GC 47\,Tucanae (NGC\,104; hereafter 47\,Tuc) is one of the nearest and most massive clusters in the Milky Way, and has long served as a benchmark for studies of stellar evolution and cluster dynamics \citep[e.g.][]{1981ApJ...245L..79C,1981ApJ...246..842F,1987PASP...99..739H,1998MNRAS.298..601C,2006ApJS..166..249M}. With a metallicity of [Fe/H]$\sim-0.75$\,dex \citep{2014ApJ...780...94C,2014A&A...572A.108T,2016MNRAS.459..610M}, it is among the most metal-rich GCs. Owing to its proximity, richness, and relatively high metallicity, 47\,Tuc has been extensively studied across a wide range of wavelengths, particularly with HST, providing key insights into its stellar populations and internal dynamics (e.g. \citealt{2009ApJ...697L..58A,2013ApJ...771L..15R}). JWST has enabled the exploration of the very low-mass regime ($M\lesssim0.1\,M_\odot$) of 47\,Tuc with unprecedented sensitivity in the near-infrared. (e.g. \citealt{2023MNRAS.521L..39N,2024ApJ...965..189M,2025A&A...694A..68S,2026A&A...706A.313V}).

In this paper, we present the first results of a series of studies based on the JWST GO-5896 programme. The focus of this work is on the characterisation of the lower MS down to the HBL, and on the corresponding LF and MF. Subsequent papers in this series will investigate the internal kinematics of the cluster and the properties of its white-dwarf population.

This paper is organised as follows. In Section\,\ref{Section2} we describe the JWST observations and data reduction. Section\,\ref{Section3} presents the PM analysis and cluster membership selection. Section\,\ref{Section4} presents the JWST colour–magnitude diagram (CMD) of 47\,Tuc and discusses the main features observed along the lower MS. Section\,\ref{Section5} describes the artificial-star tests and completeness estimates. The LF and MF are analysed in Section\,\ref{Section6}. Finally, Section\,\ref{Section7} summarises our main results.

\section{Dataset and reduction}\label{Section2}
This study is based on proprietary data from the JWST programme GO-5896\footnote{\url{https://www.stsci.edu/jwst/science-execution/program-information?id=5896}} (PI: Scalco). The programme was designed to obtain a second epoch of JWST observations of a previously imaged field in 47\,Tuc, enabling the measurement of high-precision PMs from two JWST epochs. Its primary goal was to extend the study of the internal kinematics of 47\,Tuc, in particular by probing to lower stellar masses ($\lesssim$ 0.6\,$M_{\odot}$) the recent first observational confirmation of a theoretically predicted correlation between stellar mass and rotational velocity around the cluster centre \citep{2022MNRAS.512.2584L,2023MNRAS.522L..61S}. As auxiliary science goals, the programme aims to increase the census of BD candidates in 47\,Tuc and to validate their cluster membership through high-precision JWST PMs. The new observations also allow for a deeper investigation of multiple stellar populations along the lower MS and offer new constraints on the behaviour of the white-dwarf cooling sequence of the cluster in the infrared.

The JWST programme GO-5896 was executed using the Near Infrared Camera \citep[NIRCam;][]{2023PASP..135b8001R} as the primary instrument, with the Near Infrared Imager and Slitless Spectrograph (NIRISS) observing in parallel. Figure\,\ref{FOV} shows the position of the field-of-view superimposed on a Digital Sky Survey\footnote{\url{https://archive.eso.org/dss/dss}} image of 47\,Tuc. The pointing was selected to maximise the overlap of the NIRCam field of view with the area observed during the first epoch of the JWST GO-2559\footnote{\url{https://www.stsci.edu/jwst/science-execution/program-information?id=2559}} programme (PI: Caiazzo, GO-2559 v123, represented in blue in Fig.\,\ref{FOV}) and with the deep primary field previously covered by the HST GO-11677\footnote{\url{https://archive.stsci.edu/proposal_search.php?id=11677&mission=hst}} programme (PI: Richer, represented in green in Fig.\,\ref{FOV}), while simultaneously avoiding the innermost regions of the cluster in order to minimise saturation effects. In this work, we made use exclusively of the NIRCam data; the analysis of the NIRISS parallel observations is deferred to a future, dedicated paper.

The NIRCam field of GO-5896 samples the cluster at projected distances ranging approximately from $2\,r_{\rm h}$ to $4\,r_{\rm h}$, where $r_{\rm h}=3\farcm17$ is the half-light radius of the cluster \citep{1996AJ....112.1487H,2010arXiv1012.3224H}. The observations were obtained simultaneously with the Short Wavelength (SW) and Long Wavelength (LW) channels on 8 September 2025 (epoch $\sim$2025.68). Wide-band filters were used in both channels, with F090W ($\lambda_{\rm c}\sim0.902\,\mu{\rm m},\ \Delta\lambda\sim0.194\,\mu{\rm m}$) in the SW channel and F444W ($\lambda_{\rm c}\sim4.402\,\mu{\rm m},\ \Delta\lambda\sim1.024\,\mu{\rm m}$) in the LW channel. The choice of these filters was motivated by both photometric and astrometric considerations, and by the aim of extending and complementing the wavelength coverage provided by the GO-2559 observations (obtained in F150W2 and F322W2). In particular, the F090W filter, thanks to its shorter wavelength, provides the best astrometric precision and enhances the separation between 47\,Tuc members, Small Magellanic Cloud stars, and blue point-like background galaxies. Furthermore, the F444W filter offers an optimal coverage of the reddest sources and ensures a robust astrometric second epoch for objects -- such as the brightest BDs -- that may be undetected in the bluer filters but potentially detectable in F444W. 

The six-point \texttt{FULLBOX} primary pattern was employed with a single-position \texttt{NIRCam-only} sub-pixel dither pattern. Combined with the 2$\times$1 mosaic, this resulted in a total of 12 pointings. At each pointing, a single exposure was acquired simultaneously in both channels using the \texttt{MEDIUM8} readout pattern (six groups and one integration per exposure). Each exposure consisted of multiple detector readouts: eight in the SW channel and two in the LW channel. This resulted in a total of 96 SW detector images in the F090W filter and 24 LW detector images in the F444W filter. Each detector image had an effective exposure time of 622.733\,s, corresponding to a total exposure time of 3736.396\,s ($\sim$1\,h) per pointing.

Image processing followed the procedures described in Papers I, II, and III of the ``Photometry and Astrometry with JWST'' series \citep{2022MNRAS.517..484N,2023MNRAS.525.2585N,2023AN....34430006G}, which have also been successfully applied in recent similar studies \citep{2023MNRAS.521L..39N,2024A&A...689A..59S,2025A&A...694A..68S,2025A&A...701A.169S,2024AN....34540039B,2025AN....34640125B,2025AN....34670042B}. The level-1b uncalibrated (\texttt{\_uncal}) images were first processed with a development version of the JWST pipeline\footnote{\url{https://github.com/spacetelescope/jwst}}\footnote{\url{https://zenodo.org/records/20058613}}. Pipeline stages 1 and 2 were executed to produce the level-2b (\texttt{\_cal}) images. In stage 1, all default parameters were used except for the ramp-fitting step, where the \texttt{frame\,zero} (the initial frame of each integration) was included to recover pixels saturated in the first group and extend the usable dynamic range by nearly two magnitudes. Stage 2 was executed with all default parameters.

The reduction of the \texttt{\_cal} images consisted of two main steps: first-pass and second-pass photometry. In the first pass, using the empirical library point spread functions derived by \citet{2023MNRAS.525.2585N}\footnote{\url{https://web.oapd.inaf.it/bedin/files/PAPERs_eMATERIALs/JWST/Paper_03/PSFLIB/PSF_v2.0_Aug2023/}}, we measured point spread function-fitting positions and fluxes in each individual exposure. A geometric-distortion correction was applied following \citet{2023AN....34430006G}, and the stellar positions from each image were transformed into a common reference frame using bright cluster members from \textit{Gaia} DR3 \citep{2016A&A...595A...1G,2023A&A...674A...1G} after propagating their coordinates to the epoch of the JWST observations.

Second-pass photometry was performed with a modified version of the \texttt{KS2} code \citep[see e.g.][and references therein]{2017ApJ...842....6B,2018ApJ...853...86B,2018MNRAS.481.3382N,2018ApJ...854...45L,2022LibralatoPMcat,2021MNRAS.505.3549S,2024A&A...688A.180S}. In this step, positions and fluxes were extracted using the point spread functions and transformations obtained during the first pass. \texttt{KS2} simultaneously processes all exposures, making it particularly effective for detecting and measuring faint sources not visible in individual images. In addition to positions and fluxes, \texttt{KS2} provides several diagnostic parameters, including the point spread function quality-of-fit (\texttt{QFIT}) parameter, the \texttt{RADXS} parameter \citep[a metric for comparing the shape of a source to the point spread function; see][]{2008ApJ...678.1279B,2009ApJ...697..965B}, and the local sky noise (\texttt{rmsSKY}). These diagnostics were discussed in detail by \citet{2018ApJ...853...86B,2018MNRAS.481.3382N,2021MNRAS.505.3549S}.

Photometry was calibrated to the Vega-magnitude system following \citet[][see also \citealt{2005MNRAS.357.1038B}]{2023MNRAS.525.2585N}, and the astrometry was anchored to the International Celestial Reference System using \textit{Gaia} DR3 sources in the field. We selected a sample of well-measured stars using the quality parameters provided by \texttt{KS2} (\texttt{QFIT}, \texttt{RADXS}, and \texttt{rmsSKY}). In particular, we retained sources with \texttt{QFIT} > 0 and absolute \texttt{RADXS} values smaller than 0.15 in both filters. For \texttt{rmsSKY}, we defined a magnitude-dependent threshold by tracing the typical locus of \texttt{rmsSKY} as a function of brightness and retaining sources below the 97th percentile of the \texttt{rmsSKY} distribution. The 97th percentile was adopted as a compromise between rejecting outliers and preserving sample completeness.

\begin{centering} 
\begin{figure}
 \includegraphics[width=\columnwidth]{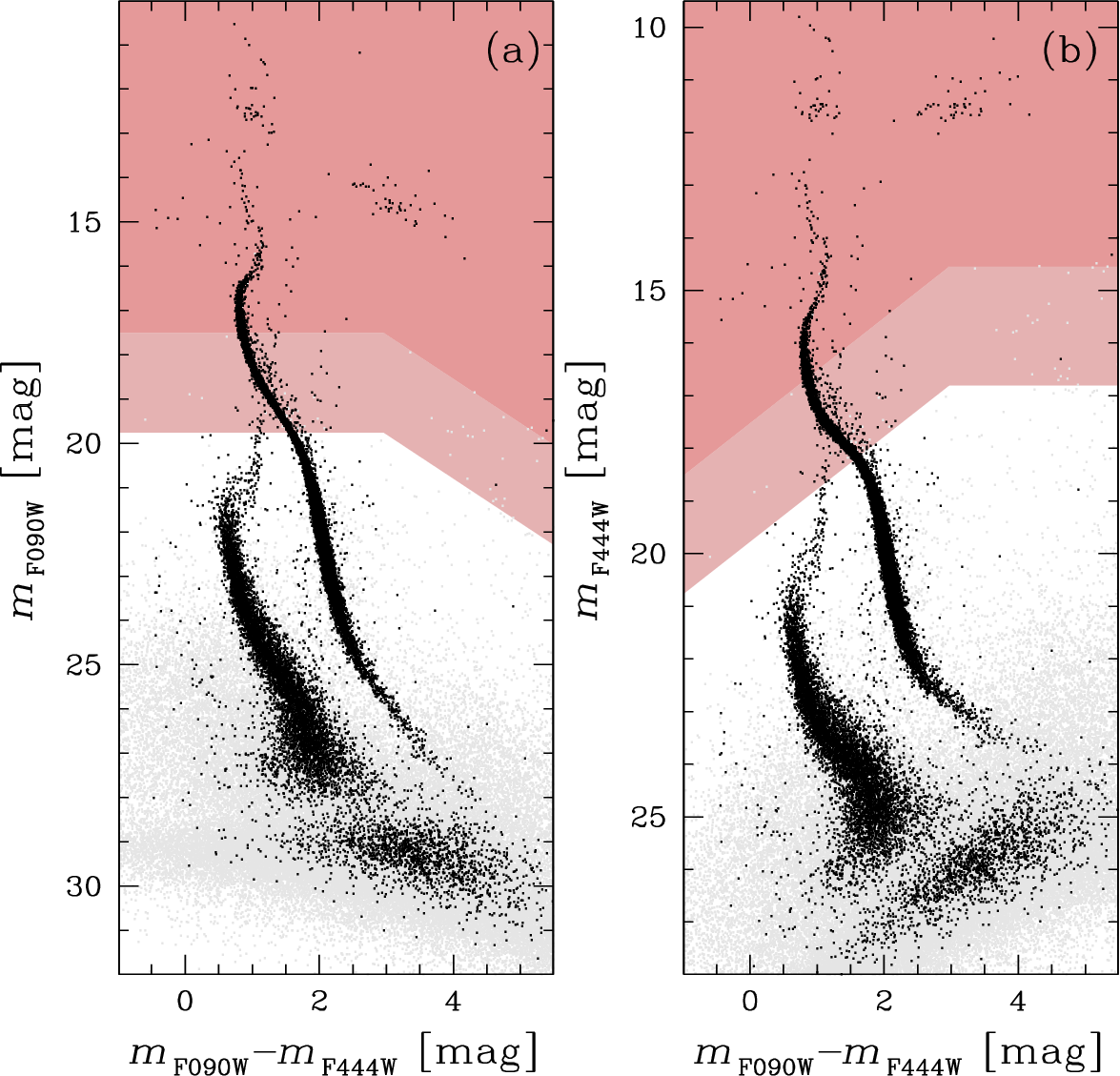}
 \caption{Colour-magnitude diagrams of 47\,Tuc. (a) $m_{\rm F090W}$ versus $m_{\rm F090W}-m_{\rm F444W}$ CMD. (b) $m_{\rm F444W}$ versus $m_{\rm F090W}-m_{\rm F444W}$ CMD. In both panels, black points indicate stars that satisfy all photometric-quality selections, while grey points mark sources that were rejected. The light-red and dark-red shaded regions highlight areas affected by saturation: light red corresponds to the regime where photometry is recovered from the \texttt{frame\,zero}, whereas dark red marks severe saturation, where even the \texttt{frame\,zero} is affected and reliable photometry cannot be obtained.} 
 \label{cmd} 
\end{figure} 
\end{centering}

Figure\,\ref{cmd} shows the $m_{\rm F090W}$ versus $m_{\rm F090W}-m_{\rm F444W}$ CMD (panel\,a) and the $m_{\rm F444W}$ versus $m_{\rm F090W}-m_{\rm F444W}$ CMD (panel\,b) for all stars in our JWST field. The selected stars delineate the cluster’s MS, its lower MS ($m_{\rm F090W}\gtrsim20$\,mag and $m_{\rm F444W}\gtrsim18$\,mag), and the white dwarf cooling sequence in the lower-left region of the diagrams. In addition, a MS belonging to the Small Magellanic Cloud is visible between the 47\,Tuc MS and its white dwarf cooling sequence.

In the severe saturation regime, the photometry becomes unreliable and non-physical features appear in the sequences. Stars in the saturated regions must therefore be treated with caution, and none of them is used in the scientific analysis presented in this work. 

\section{Proper motions}\label{Section3}
To obtain the PMs, we made use of NIRCam data from the first epoch of the JWST GO-2559 programme (visit v123; epoch $\sim$2022.7), the second epoch of the same programme (visit v51; epoch $\sim$2023.3), and the second visit of the JWST GO-2560\footnote{\url{https://www.stsci.edu/jwst/science-execution/program-information?id=2560}} programme (PI: Marino, visit v2; epoch $\sim$2023.7), as presented by \citet{2025A&A...694A..68S}. Together with the GO-5896 observations, these datasets provide a temporal baseline of $\sim$3 years for the PM measurements. Using this combination of epochs, we were able to measure reliable PMs over approximately 75\% of the area covered by the GO-5896 NIRCam observations. A complete description of the GO-2559 and GO-2560 datasets, as well as of their reduction procedures, was provided by \citet{2025A&A...694A..68S}.

The PMs were computed using the method described by \citet{2021MNRAS.505.3549S} and introduced by \citet[][see also \citealt{2018ApJ...853...86B,2018ApJ...854...45L,2022LibralatoPMcat}]{2014ApJ...797..115B}. This iterative method treats each exposure as an independent epoch. The procedure involves two main steps: first, stellar positions from individual images are transformed into a common reference frame using a six-parameter linear transformation; second, these transformed positions are fitted as a function of time with a least-squares linear regression. The slope of this fit, obtained after multiple rounds of outlier rejection, provides the PM measurement. For this analysis, we relied exclusively on the SW-channel exposures, which provide significantly higher astrometric precision than their LW counterparts. To account for high-frequency systematic variations, we applied the correction described by \citet{2021MNRAS.505.3549S}, subtracting the median PM of the 100 nearest likely cluster members (excluding the target star itself).

\begin{centering} 
\begin{figure*}
 \includegraphics[width=\textwidth]{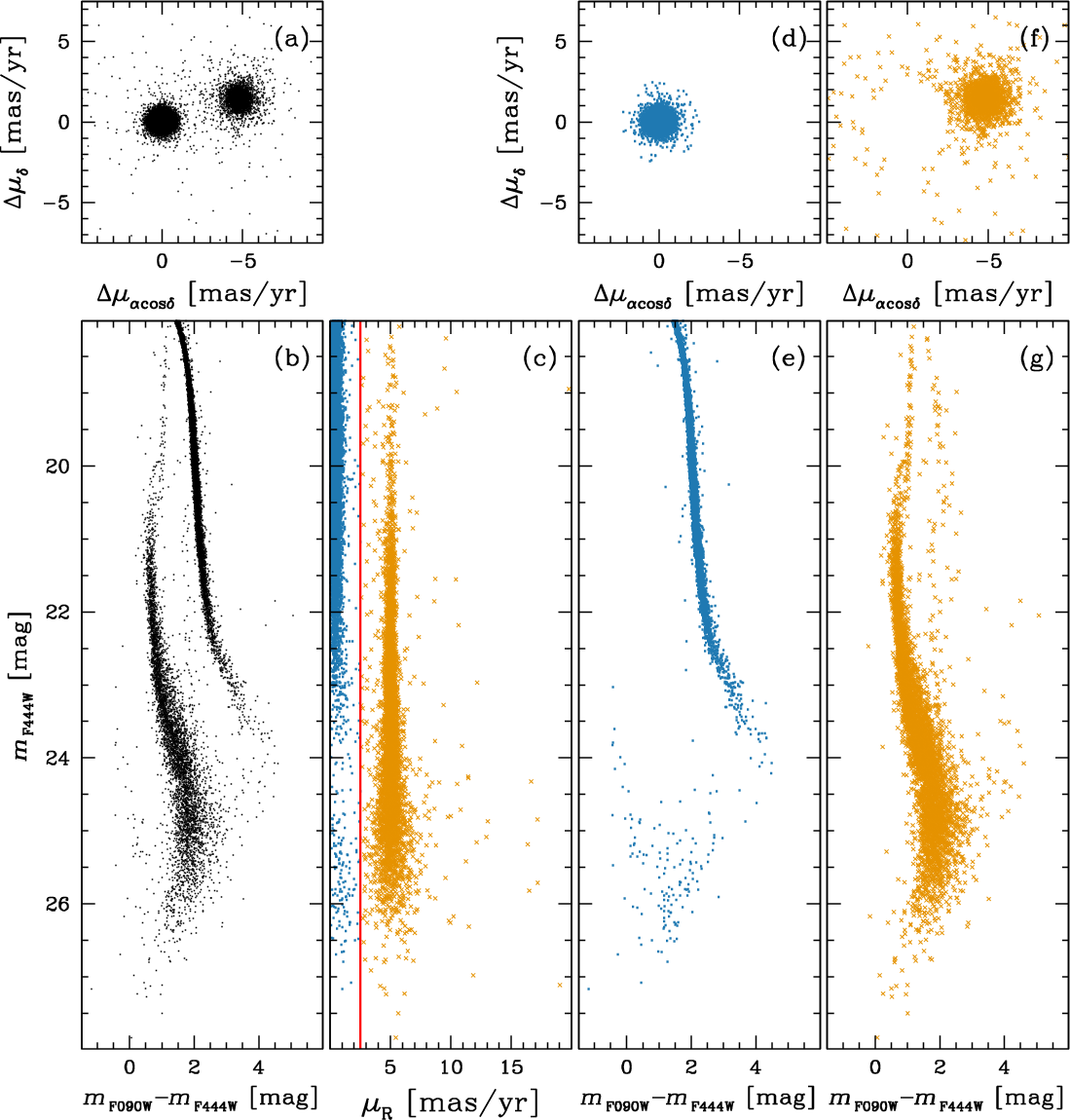}
 \caption{Cluster membership selection based on the obtained PMs. Only sources that satisfy the photometric-quality criteria and have measurable PMs are shown. (a)-(b) Vector-point diagram and $m_{\rm F444W}$ versus $m_{\rm F090W}-m_{\rm F444W}$ CMD, respectively. (c) $m_{\rm F444W}$ magnitude versus the one-dimensional PM ($\mu_{\rm R}$). The red line separates cluster members from Small Magellanic Cloud and field stars. (d)-(e) Vector-point diagram and $m_{\rm F444W}$ versus $m_{\rm F090W}-m_{\rm F444W}$ CMD for stars that passed the PM selection. (f)-(g) same diagrams for stars that did not pass the PM selection. In panels\,(c) through (g), sources that passed the PM selection are represented by blue dots, while those that did not pass are depicted as orange crosses.} 
 \label{pm} 
\end{figure*} 
\end{centering}

The resulting PMs are shown in Fig.\,\ref{pm}, which includes only stars that satisfy the photometric-quality criteria described in the previous section and for which a reliable PM measurement could be obtained. Panels\,(a) and (b) show the vector-point diagram and the $m_{\rm F444W}$ versus $m_{\rm F090W}-m_{\rm F444W}$ CMD, respectively. Since PMs are measured relative to the mean motion of the cluster, 47\,Tuc members are centred around $(0,0)\,\mathrm{mas\,yr^{-1}}$) in the vector-point diagram. A second, well-separated concentration of points, visible near $(5,1.5)\,\mathrm{mas\,yr^{-1}}$, corresponds to stars belonging to the Small Magellanic Cloud.

Panel\,(c) presents the one-dimensional PM ($\mu_{\rm R}$, computed as the quadrature sum of the two PM components) as a function of $m_{\rm F444W}$. Cluster members form a narrow sequence at $\mu_{\rm R} \lesssim 1$\,mas\,yr$^{-1}$, while non-members extend to significantly larger values, with the bulk of Small Magellanic Cloud stars peaking around $\mu_{\rm R} \sim 5$\,mas\,yr$^{-1}$.

The improvement relative to the PMs obtained by \citet{2025A&A...694A..68S} is substantial (see panel\,(d) of their Fig.\,3). In that study, the temporal baseline was limited to only 7\,months, whereas the $\sim$3-year baseline used here greatly enhances the astrometric precision. As a result, cluster and non-cluster populations remain cleanly separated down to the end of the MS, enabling reliable membership classification even at the faintest magnitudes accessible in our data.

We defined a conservative PM selection criterion to separate cluster members from Small Magellanic Cloud stars and field objects, indicated by the vertical red line at 2.5\,mas\,yr$^{-1}$. Panels\,(d) and (e) show the vector-point diagram and $m_{\rm F444W}$ versus $m_{\rm F090W}-m_{\rm F444W}$ CMD for the stars that passed the PM selection, while panels\,(f) and (g) show the same diagrams for stars that did not pass the PM selection.

\begin{centering} 
\begin{figure}
 \includegraphics[width=\columnwidth]{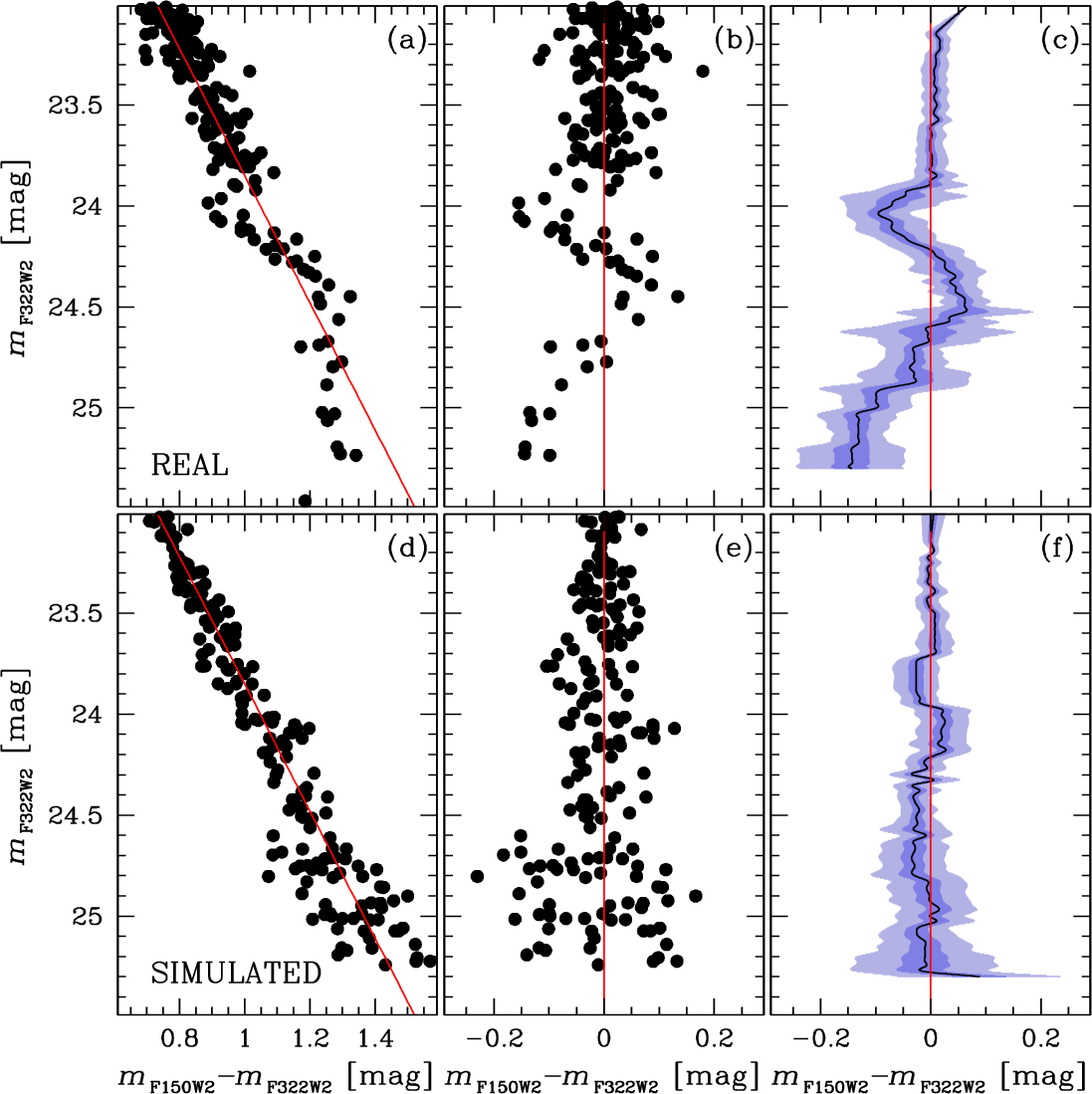}
 \caption{Statistical significance of the kink in the lower MS of 47\,Tuc. Panels\,(a)-(c) show the observed data, while panels\,(d)-(f) show a synthetic stellar sample. Panels\,(a) and (d) show the $m_{\rm F322W2}$ versus $m_{\rm F150W2}-m_{\rm F322W2}$ CMDs. The red line represents the adopted fiducial. Panels\,(b) and (e) show the corresponding CMDs after verticalization with respect to the fiducial. Panels\,(c) and (f) show the median of the verticalized colour as a function of $m_{\rm F322W2}$ (black line), obtained from a set of shifted magnitude binnings. The light- and dark-blue shaded regions indicate the corresponding 1$\sigma$ and 3$\sigma$ confidence intervals.}
\label{kink1} 
\end{figure} 
\end{centering}

From panel\,(e) of Fig.\,\ref{pm} the MS can be reliably traced down to $m_{\rm F444W} \sim 24.25$\,mag. We note that the use of SW astrometry to evaluate the PMs can hinder the detection of the reddest and faintest sources, which may not be visible in the F090W filter but are detected in F444W. This may effectively limit the depth of the stellar sample that we are recovering. To investigate this limitation, we performed an alternative PM analysis to probe the lower part of the MS. In particular, we evaluated the PMs by relying only on the astrometry derived from the two reddest available filters, namely the F444W data from our GO-5896 observations and the F322W2 data from the deepest available epoch, GO-2559 v123.

The resulting PMs are shown in Fig.\,\ref{pm2} of Appendix\,\ref{Appendix:A}, which is analogous to Fig.\,\ref{pm} but includes a CMD constructed using only the F322W2 and F444W filters. To obtain the PMs, we performed the source finding independently for each epoch, using only the F322W2 and F444W images for the finding. Proper motions were then estimated by measuring the displacement of sources between the two epochs. Photometric selections were applied only in F322W2 and F444W, thus allowing the inclusion of very red sources not detected in F090W. We note that the spatial coverage of this analysis is slightly smaller than that of Fig.\,\ref{pm}, as it is limited to the overlap between the GO-5896 and GO-2559 v123 datasets. As a result, the number of sources is correspondingly reduced.

As expected, this approach yields significantly larger PM uncertainties than those obtained in the previous analysis. This is primarily due to the use of LW astrometry, which provides lower astrometric precision compared to the SW data adopted above. We applied the same PM selection criteria employed in Fig.\,\ref{pm}. As shown in Fig.\,\ref{pm2}, the increased PM errors prevent a significant improvement in the faint limit and do not allow us to probe substantially deeper than in the previous analysis, with the MS extending only down to $m_{\rm F444W} \sim 23.75$\,mag. In the following analysis, we restrict our sample to the stars shown in panel\,(e) of Fig.\,\ref{pm}, that is, to those satisfying both the photometric-quality criteria and the PM-selection criterion, resulting in a final sample of $\approx$10,000 cluster members.

\begin{centering} 
\begin{figure*}
 \includegraphics[width=\textwidth]{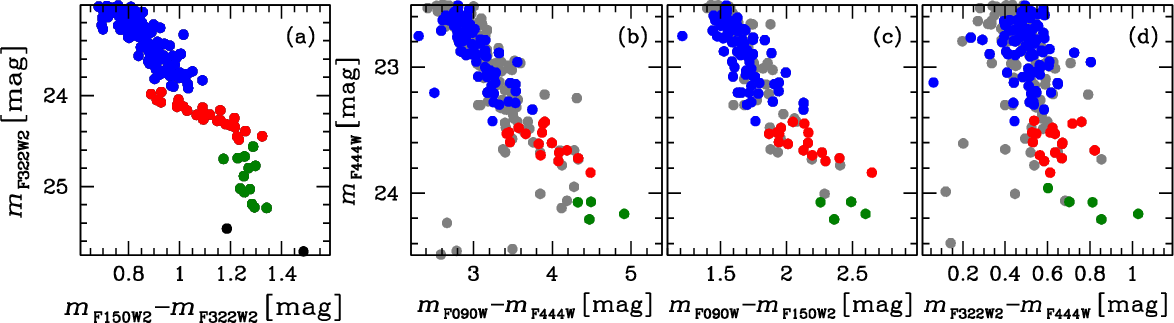}
 \caption{Kink in the lower MS of 47 Tuc. (a) $m_{\rm F322W2}$ versus $m_{\rm F150W2}-m_{\rm F322W2}$ CMD of the lower MS of 47\,Tuc. Stars associated with the three segments of the kink are highlighted with different colours (blue, red, and green). The same stars are shown in three additional CMDs adopting the same colour coding: (b) $m_{\rm F444W}$ versus $(m_{\rm F090W}-m_{\rm F444W})$, (c) $m_{\rm F444W}$ versus $(m_{\rm F090W}-m_{\rm F150W2})$, and (d) $m_{\rm F444W}$ versus $(m_{\rm F322W2}-m_{\rm F444W})$.}
 \label{kink2} 
\end{figure*} 
\end{centering}

\section{Results and discussion}

\subsection{The colour magnitude diagram of 47\,Tuc}\label{Section4}

One of the most remarkable features revealed by JWST observations of 47\,Tuc is the complex morphology of the lower MS. In particular, recent JWST studies have reported the presence of a discontinuity in the slope of the MS at the very low-mass end of the sequence, close to the HBL \citep{2024ApJ...965..189M,2025A&A...694A..68S}. This feature, referred by \citet{2025A&A...694A..68S} to as the ``kink'', appears as a change in the morphology of the MS and is accompanied by a small gap and a change in the colour-magnitude slope of the sequence. This feature is thought to be associated with changes in the atmospheric chemistry of very low-mass stars, in particular the transition in carbon chemistry from predominantly oxidised to reduced (methane-dominated) equilibrium \citep{CO_ratio}. As discussed in \citet{2025A&A...694A..68S}, this transition in 47\,Tuc appears to occur at significantly earlier spectral types than in the field population \citep{methane_in_field_BDs}, suggesting a lower carbon-to-oxygen ratio in lower MS members compared to their higher-mass counterparts.

To illustrate the morphology of the kink, panel\,(a) of Fig.\,\ref{kink1} shows the lower MS of 47\,Tuc in the $m_{\rm F322W2}$ versus $m_{\rm F150W2}-m_{\rm F322W2}$ CMD, where the feature was previously investigated by \citet{2025A&A...694A..68S}. The red line represents a least-squares fit to the observed lower MS. The sequence remains straight, well defined, and follows the fitted relation down to $m_{\rm F322W2}\sim23.9$\,mag. At fainter magnitudes, however, it begins to display a series of discontinuities and departs from the extrapolated trend. In particular, stars in the interval $23.9\lesssim m_{\rm F322W2}\lesssim24.25$\,mag are systematically located on the blue side of the red line, whereas stars in the range $24.25\lesssim m_{\rm F322W2}\lesssim24.7$\,mag are systematically located on the red side. At still fainter magnitudes ($24.7\lesssim m_{\rm F322W2}\lesssim25.4$\,mag), the stars are again systematically located on the blue side of the red line.

To quantify the significance of these deviations, we performed a test to assess whether the observed sequence is consistent with a smooth, non-discontinuous trend. The procedure is illustrated in Fig.\,\ref{kink1}. We adopted the least-squares fit shown in panel\,(a) as our reference model in the absence of discontinuities. Although a linear relation may not be the most appropriate description of the lower MS over a broad magnitude range, the narrow interval considered here makes a linear approximation reasonable. We first verticalized the CMD with respect to the least-squares fit line; the resulting verticalized CMD is shown in panel\,(b). We then divided the sequence into magnitude bins of width 0.25\,mag and measured, for each bin, the median of the verticalized colour and the corresponding uncertainty, defined as the 68th percentile of the colour distribution divided by $\sqrt{N-1}$, where $N$ is the number of stars in the bin. To minimise any dependence on the arbitrary choice of the bin boundaries, we repeated the analysis using a series of ten magnitude bins shifted by one tenth of the bin width (0.025\,mag). The measurements obtained from all shifted binnings were then combined to derive a continuous estimate of the median colour of the sequence. The resulting trend is shown in panel\,(c) as a black line, while the light- and dark-blue shaded regions indicate the corresponding 1$\sigma$ and 3$\sigma$ confidence intervals. As can be seen from panel\,(c), the discontinuities deviate from zero by more than 3$\sigma$.

To assess whether the observed behaviour could be explained by photometric scatter alone, we compared the observed distribution with that obtained from a synthetic stellar sample. The colours of the simulated stars were randomly drawn from Gaussian distributions centred on the least-squares fiducial and having dispersions equal to the colour uncertainties derived from the artificial-star tests at the corresponding magnitudes. The number of simulated stars was chosen to match that of the observed sample. The resulting CMD is shown in panel\,(d). The same verticalization procedure was then applied to the simulated sample, yielding the verticalized CMD shown in panel\,(e). We applied to the verticalised sequence the same binning, median, and uncertainty-estimation procedure adopted for the observed sample. The resulting median colours and their corresponding uncertainties are shown in panel\,(f). In contrast to the observed sample, the median colour of the simulated sequence remains consistent with zero over the entire magnitude range. While these results support the presence of the observed discontinuities, we note that the number of stars involved remains limited. Additional observations and larger stellar samples will be required for a more robust assessment of their statistical significance.

To investigate whether the kink can be identified in the F090W and F444W photometry analysed in this work, we divided the stars highlighted in panel\,(a) of Fig.\,\ref{kink1} into three groups, corresponding to different segments of the sequence. The three groups are shown in panel\,(a) of Fig\,\ref{kink2} in green, red and blue. Panels\,(b), (c) and (d) of Fig.\,\ref{kink2} show these same stars, using the same colour coding, in the $m_{\rm F444W}$ versus $m_{\rm F090W}-m_{\rm F444W}$, $m_{\rm F444W}$ versus $m_{\rm F090W}-m_{\rm F150W2}$, and $m_{\rm F444W}$ versus $m_{\rm F322W2}-m_{\rm F444W}$ CMDs, respectively. Small changes in the morphology of the sequence can be seen at the transition between the three groups, particularly in panels\,(b) and (c). However, these features are considerably weaker than those observed in panel\,(a) and involve only a small number of stars, making any interpretation uncertain.

The weak appearance of the feature in these colour combinations is qualitatively consistent with its proposed interpretation as a consequence of changes in atmospheric chemistry, in particular the transition in carbon chemistry toward a methane-dominated equilibrium. Its photometric impact is expected to be reduced in these filters, which are less sensitive to the molecular absorption features associated with this transition \citep{2025A&A...694A..68S}.

Panel (a) of Fig.\,\ref{iso} shows the $m_{\rm F444W}$ versus $m_{\rm F090W}-m_{\rm F444W}$ CMD of our selected sample of stars. We overplotted on the CMD the theoretical isochrones, computed in \citet{2024ApJ...961..139G} for the age and chemical composition of 47\,Tuc. The three models shown in the figure --  \texttt{G24B}, \texttt{G24RL}, and \texttt{G24R} -- differ in their adopted oxygen abundances ($[\mathrm{O/Fe}]=0.48$\,dex for \texttt{G24B}, $[\mathrm{O/Fe}]=0.33$\,dex for \texttt{G24RL} and $[\mathrm{O/Fe}]=0.03$\,dex for \texttt{G24R}), and they represent the best fits to the blue tail, ridgeline and red tail of the observed scatter in the HST Wide Field Camera 3 (WFC3) $m_{\rm F110W}-m_{\rm F160W}$ colour of the lower MS from \citet{2012AJ....143...11K}. Synthetic photometry in the $m_{\rm F090W}-m_{\rm F444W}$ CMD space was calculated using \texttt{BasicATLAS} \citep{BasicATLAS}, adopting the optical reddening of $E(B-V)=0.04$\,mag \citep{2024ApJ...961..139G}, distance modulus of $13.24$\,mag \citep{distance_modulus}, total-to-selective extinction of $R_V=3.1$ and the reddening law from \citet{G24_law_0,G24_law_software} (also see \citealt{G24_law_1,G24_law_3,G24_law_2,G24_law_4}). Note that these models do not account for the mass dependence of $[\mathrm{O/Fe}]$, hypothesised in \citet{2025A&A...694A..68S}. For comparison, panel\,(b) shows the $m_{\rm F322W2}$ versus $m_{\rm F150W2}-m_{\rm F322W2}$ CMD and \texttt{G24} isochrones from \citet{2025A&A...694A..68S}, together with recent models from \citet{2026A&A...706A.313V}.

The three isochrones closely overlap in the upper part of the CMD, where the MS runs almost vertically. At fainter magnitudes, however, the models progressively diverge. As the MS begins to bend toward redder colours, the \texttt{G24B} and \texttt{G24RL} isochrones shift toward bluer colours, while the \texttt{G24R} model more closely follows the observed locus of the lower MS. Although none of the available models provides a satisfactory quantitative fit to the faintest portion of the sequence, \texttt{G24R} is the least discrepant among those considered here and provides the closest overall match to the morphology of the lower MS. As discussed in \citet{2025A&A...694A..68S}, the remaining discrepancies may indicate oxygen abundances lower than those adopted in the \texttt{G24R} model or shortcomings in the current treatment of molecular opacities. In that work, dedicated atmosphere calculations were performed by varying O, C, and Ti abundances. The analysis showed that reproducing the morphology of the faintest portion of the MS requires conditions favouring enhanced CH$_4$ formation at low effective temperatures and identified the atmospheric C/O ratio as the key parameter controlling this behaviour. Within the explored parameter space, these conditions were most naturally obtained for oxygen-poor compositions. Notably, the \texttt{G24R} isochrone displays a subtle discontinuity in the slope of the sequence around $m_{\rm F444W}\sim23.7$\,mag, providing an approximate indication of the expected location of the kink.

Selected stellar masses are indicated along the isochrones. The HBL ($\sim0.074\,M_\odot$), as estimated by \citet{2024ApJ...961..139G}, is also shown and lies below the faintest members in our sample, which reach masses of approximately $\sim0.077\,M_\odot$. In the next section, we provide an estimate of the completeness of our data.

\subsection{Artificial stars}\label{Section5}

\subsubsection{Artificial star tests and completeness}

We estimated the completeness of our catalogue using artificial stars. To incorporate the PM selection into the completeness estimate, we generated the artificial stars independently in each JWST epoch dataset and estimated their PMs following the same procedure adopted for the real stars. Because artificial stars are injected at identical positions in all epochs, any measured PM reflects only noise-induced offsets (e.g. point spread function-modelling uncertainties, transformation residuals, cosmic rays, detector artefacts).

\begin{centering} 
\begin{figure*}
 \includegraphics[width=\textwidth]{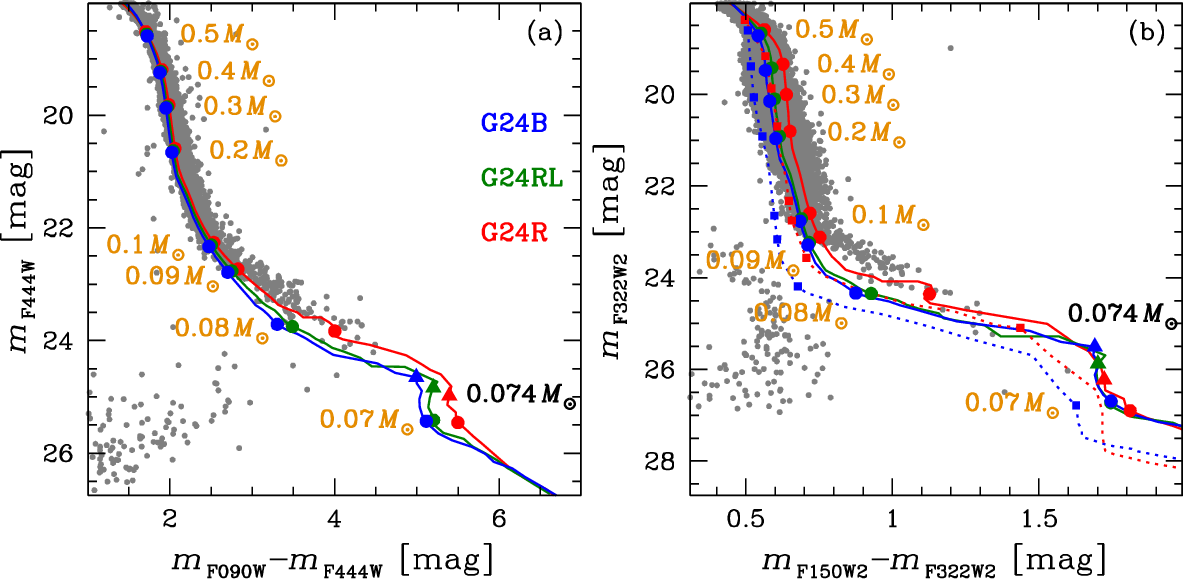}
 \caption{Comparison between the observed CMDs and theoretical isochrones. (a) $m_{\rm F444W}$ versus $m_{\rm F090W}-m_{\rm F444W}$ CMD. (b) $m_{\rm F322W2}$ versus $m_{\rm F150W2}-m_{\rm F322W2}$ CMD. In both panels, the blue, green, and red solid curves show the \texttt{G24B}, \texttt{G24RL}, and \texttt{G24R} theoretical isochrones, respectively. In panel (b), the dotted curves show the models from \citet{2026A&A...706A.313V}. Selected stellar masses are labelled along the isochrones and indicated by circles for the \texttt{G24B}, \texttt{G24RL}, and \texttt{G24R} models, and by squares for the \citet{2026A&A...706A.313V} models. The triangles mark the position of the HBL at $\sim0.074\,M_\odot$, for the \texttt{G24B}, \texttt{G24RL}, and \texttt{G24R} isochrones.}
 \label{iso} 
\end{figure*} 
\end{centering}

To assign magnitudes to the artificial stars, we defined a fiducial line tracing the observed MS. The fiducial was defined in multiple CMDs in order to cover all the filters available in the different JWST datasets used to compute the PMs and to assign consistent magnitudes to the artificial stars in each band. The fiducial was defined manually by tracing the observed MS down to the faintest magnitudes where the sequence can be reliably identified, and then extending it toward fainter magnitudes using the theoretical isochrones presented in the previous section. We generated $10^5$ artificial stars uniformly distributed across the region of the field where multi-epoch JWST observations are available and PMs could be reliably measured. The $m_{\rm F444W}$ magnitudes were drawn uniformly in the range $17.5 < m_{\rm F444W} < 28.5$\,mag, while the corresponding magnitudes in the other filters were assigned according to the adopted fiducial line.

The artificial stars were injected, detected, and measured using KS2, following exactly the same procedures adopted for the real stars. An artificial star was considered recovered if (i) its input and output positions differed by less than one pixel; (ii) its input and output magnitudes differed by less than 0.75\,mag (corresponding to $\sim$2.5\,log\,2) for each filter; and (iii) it passed the same photometric-quality selections applied to real sources. In addition, the recovered artificial star was required to satisfy the same PM-selection criterion used for real stars (see panel\,(c) of Fig.\,\ref{pm}). This allows us to quantify the fraction of stars lost due to PM-selection effects and to include this correction in the final completeness estimates.

\begin{centering} 
\begin{figure*}
 \includegraphics[width=\textwidth]{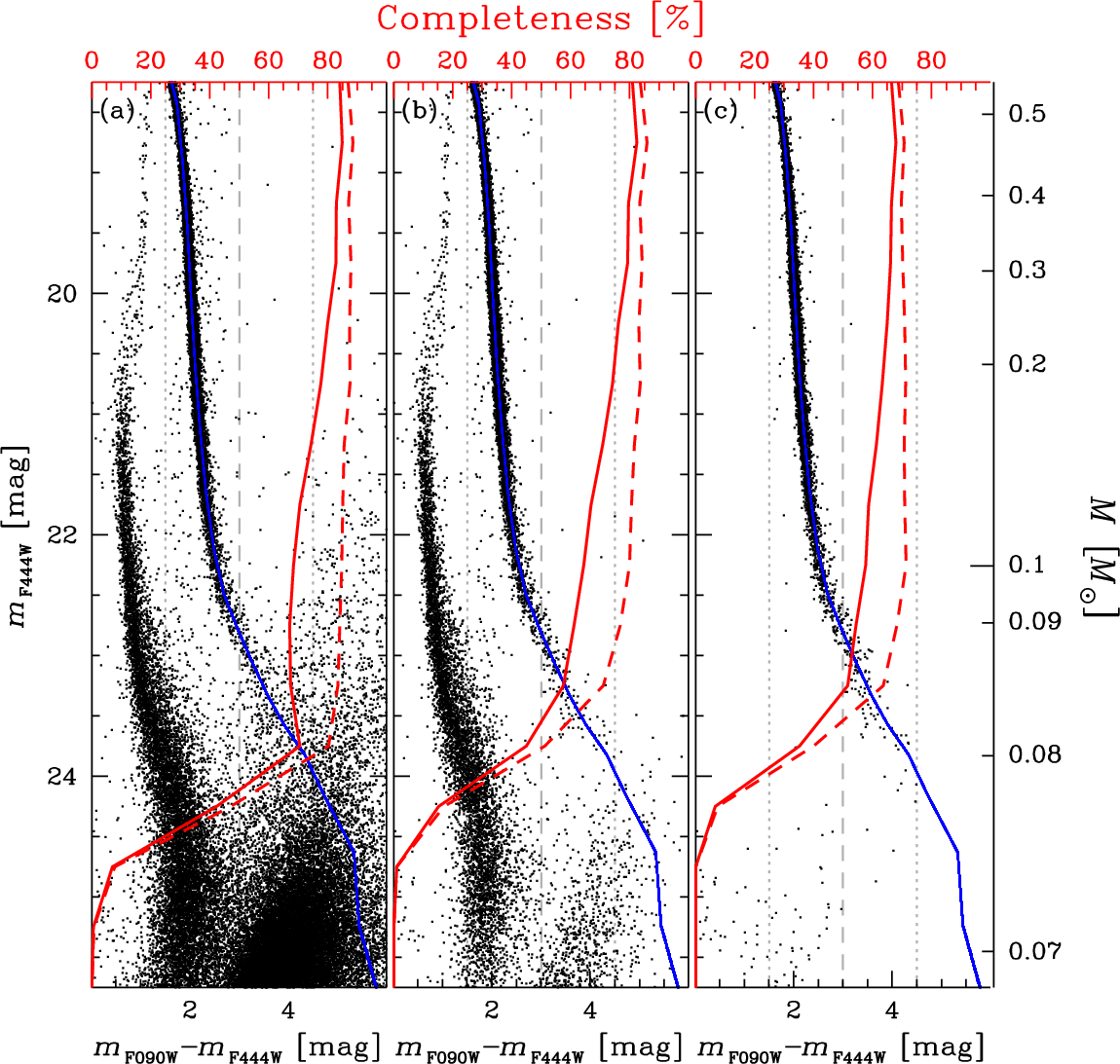}
 \caption{Completeness estimated using artificial stars. The three panels show the $m_{\rm F444W}$ versus $m_{\rm F090W}-m_{\rm F444W}$ CMD for (a) the catalogue without applying \texttt{RADXS} and PM selections, (b) after applying the \texttt{RADXS} selection but not the PM selection, and (c) after applying both \texttt{RADXS} and PM selections. The blue curve represents the fiducial used to inject the artificial stars. The red solid curve shows the completeness as a function of magnitude for each CMD and corresponding set of selections, while the red dashed curve indicates $c_g$, the completeness measured in the dark usable areas of the detectors where faint sources can be reliably detected. The upper horizontal axis reports the completeness values. The right vertical axis shows the stellar masses estimated using the mass-luminosity relation of the \texttt{G24R} model. The vertical dashed and dotted lines indicate completeness levels of 25\%, 50\%, and 75\%.} 
 \label{comp} 
\end{figure*} 
\end{centering}

Figure\,\ref{comp} illustrates the completeness of our catalogue under different selection criteria. The three panels show the $m_{\rm F444W}$ versus $m_{\rm F090W}-m_{\rm F444W}$ CMD for (a) the catalogue without applying either \texttt{RADXS} or PM selections, (b) after applying the \texttt{RADXS} selection but not the PM selection, and (c) after applying both \texttt{RADXS} and PM selections. The blue line shows the fiducial used to inject the artificial stars. The red solid curve represents the completeness as a function of magnitude for each CMD and corresponding set of selections, while the red dashed curve shows $c_g$, the completeness measured in the dark usable areas of the detectors where faint sources can be reliably detected \citep[see][for details]{2008ApJ...678.1279B,2009ApJ...697..965B}. The right-hand vertical axis indicates the stellar masses estimated using the mass-luminosity relation of the \texttt{G24R} model. The completeness progressively decreases as increasingly stringent selection criteria are applied. We note that the completeness does not reach 100\% even at bright magnitudes in any of the cases shown in Fig.\,\ref{comp}, as the recovery, photometric-quality, and PM-selection criteria adopted in this work lead to the rejection of even some of the brightest sources. In panel\,(a), where neither \texttt{RADXS} nor PM selections are applied, the completeness remains above $\sim50\%$ down to $m_{\rm F444W}\sim24$\,mag ($\sim0.079\,M_\odot$) and drops to $\sim0\%$ around $m_{\rm F444W}\sim25.5$\,mag ($\sim0.069\,M_\odot$). After applying the \texttt{RADXS} selection (panel\,b), the completeness remains above $\sim50\%$ down to $m_{\rm F444W}\sim23.5$\,mag ($\sim0.082\,M_\odot$) and reaches $\sim0\%$ around $m_{\rm F444W}\sim25$\,mag ($\sim0.074\,M_\odot$). Finally, after applying both the \texttt{RADXS} and PM selections (panel\,c), the completeness remains above $\sim50\%$ down to $m_{\rm F444W}\sim23.25$\,mag ($\sim0.084\,M_\odot$) and drops to zero at $m_{\rm F444W}\sim24.75$\,mag ($\sim0.075\,M_\odot$, i.e. just above the estimated HBL for 47\,Tuc.) This behaviour is also reflected in the decreasing number of sources visible in the CMD, particularly at its faint end. In particular, the PM selection is crucial for removing spurious detections that populate the lower part of the CMD and contaminate the MS at its faint end.

\subsubsection{Comparison to theoretical-completeness limit}

The BD cooling sequence in 47\,Tuc has been reported in several recent studies, beginning with \citet{2023MNRAS.521L..39N}. However, in most cases, the bona fide nature of these candidates could not be established due to the lack of reliable PM-based membership constraints (see further discussion of this issue in Section~\ref{sec:faint_end_LF}). Given that our dataset provides the deepest PM-validated photometry of 47\,Tuc to date, it is instructive to reassess the detectability of BDs in the cluster. In particular, we compare both empirically measured and theoretically predicted completeness limits to the expected location of the HBL, in order to evaluate whether such detections are feasible in presently available JWST photometry.

In this subsection, we compare the completeness derived from our artificial star tests with the theoretical predictions of \citet{Roman_GC_ages}. The latter are based on the expected JWST performance from \citet{JWST_ETC}, assuming two epochs of observations in the F150W2 and F322W2 bands separated by several years. In those simulations, a source is considered recovered if its $m_{\rm F150W2}-m_{\rm F322W2}$ colour uncertainty is $<0.25\ \mathrm{mag}$, and its PM is consistent with 47\,Tuc membership at $\geq 95\%$ confidence.

\begin{centering} 
\begin{figure}
 \includegraphics[width=\columnwidth]{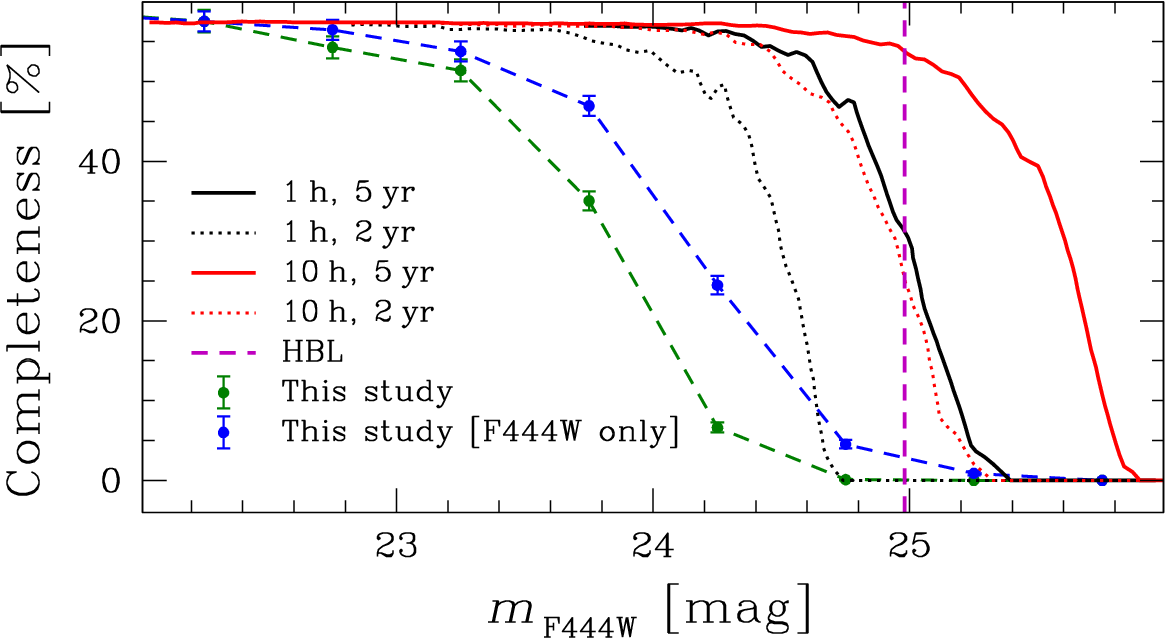}
 \caption{Measured and theoretical completeness of membership-verified JWST photometry of 47\,Tuc as a function of the $m_{\rm F444W}$ magnitude. Theoretical completeness (solid and dotted black and red curves) are adopted from \citet{Roman_GC_ages} for four distinct observing strategies, defined by exposure times and temporal baselines. Theoretical completeness curves have been converted to $m_{\rm F444W}$ using the \texttt{G24R} isochrone. Green markers show the empirical completeness calculated in this study using artificial star tests. Blue markers show an alternative completeness curve that waives the requirement of reliable $m_{\rm F090W}$ photometry. All completeness curves were scaled to the empirical value at $m_{\rm F444W}=22$\,mag. The estimated magnitude of the HBL is shown with a vertical magenta line.} 
 \label{completeness_vs_predicted} 
\end{figure} 
\end{centering}

Figure~\ref{completeness_vs_predicted} compares the theoretical completeness from \citet{Roman_GC_ages} to our artificial star-based completeness estimates. The theoretical curves correspond to four observing strategies (two $1\ \mathrm{h}$ or $10\ \mathrm{h}$ exposures separated by $2$ or $5$ years), while our measurements are shown as green markers. As expected, our artificial star-based completeness is systematically lower than the theoretical predictions for three main reasons: \textit{(1)} we use narrower F090W and F444W filters instead of the ultra-wide F150W2 and F322W2 bands adopted in \citet{Roman_GC_ages}; \textit{(2)} the F090W band is too blue for efficient detection of low MS stars and BDs, and \textit{(3)} the simulations in \citet{Roman_GC_ages} ignore crowded photometry effects. To isolate the impact of \textit{(2)}, we recomputed the artificial star-based completeness without the F090W band and plotted the result in Fig.\,\ref{completeness_vs_predicted} with blue markers.

Although none of the simulated observing strategies in Fig.\,\ref{completeness_vs_predicted} exactly match our dataset (which consists of $\sim4\ \mathrm{h}$ and $\sim1\ \mathrm{h}$ exposures separated by $\sim3$ years), our completeness should approximately fall between the predictions for two $1\ \mathrm{h}$ exposures separated by $2$ and $5$ years (dotted and solid black curves in Fig.\,\ref{completeness_vs_predicted}, respectively). Indeed, the artificial star-based faint limit for the F444W-only case (blue markers) is found within the expected range. However, requiring reliable F090W photometry shifts the faint limit brighter by $\sim1\ \mathrm{mag}$ (green markers), and both the F444W-only and total completeness show a more gradual decline compared to the models.

Overall, these results indicate that current JWST observations of 47\,Tuc are not sufficiently deep to reliably trace the cluster’s BD cooling sequence, rendering earlier claims of its detection uncertain. Nevertheless, the agreement in overall performance suggests that an additional epoch of longer-wavelength, ultra-wide-band observations should be sufficient to enable a robust detection of the brightest BDs and of the upper part of the cluster's BD cooling sequence.

\subsection{Luminosity and mass function}\label{Section6}

In this section, we present the LF and MF of 47\,Tuc along the MS and low-MS using our JWST catalogue.

\subsubsection{Observed luminosity and mass functions}\label{obs_lum_MF}

\begin{centering} 
\begin{figure}
 \includegraphics[width=\columnwidth]{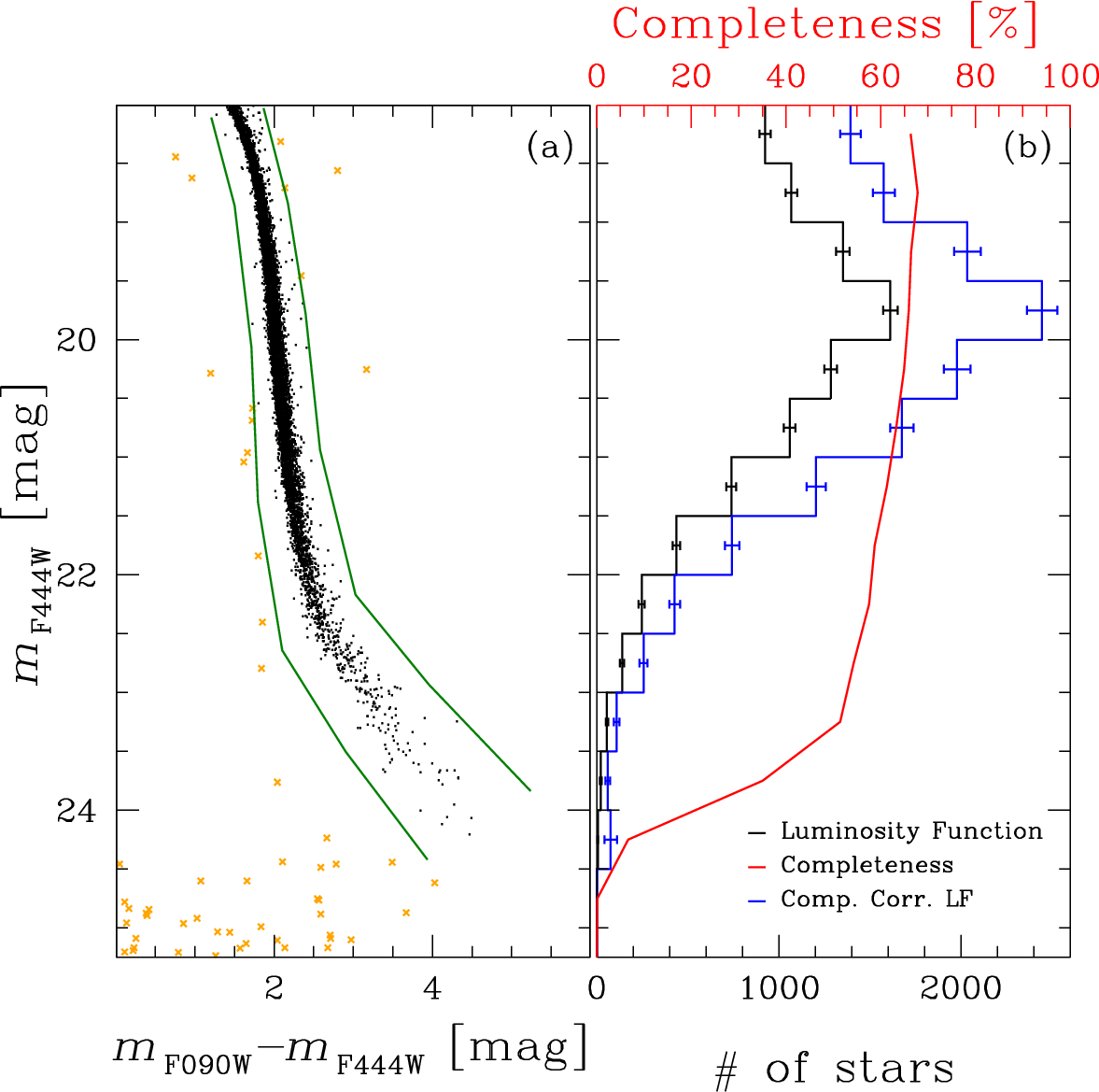}
 \caption{Luminosity function of the MS and lower MS of 47\,Tuc. (a) $m_{\rm F444W}$ versus $m_{\rm F090W}-m_{\rm F444W}$ CMD of the selected sample. The green lines mark the fiducial boundaries adopted to define the cluster MS. Stars lying between the two fiducials are shown as black points, while stars outside this region are shown as orange crosses. (b) Luminosity function as a function of $m_{\rm F444W}$. The black histogram shows the raw star counts in bins of 0.5\,mag, while the blue histogram shows the counts corrected for completeness. The completeness function is overplotted as a red curve, with the corresponding values indicated on the upper horizontal axis. Error bars on the black histogram represent Poisson uncertainties, while those on the completeness-corrected LF include Poisson errors propagated through the completeness correction.} 
 \label{lf} 
\end{figure} 
\end{centering}

Figure\,\ref{lf} presents the LF of the MS and lower MS of 47\,Tuc, derived from the dataset presented in this work. Panel\,(a) shows the $m_{\rm F444W}$ versus $m_{\rm F090W}-m_{\rm F444W}$ CMD of the selected stellar sample, from the saturation limit ($m_{\rm F444W}\sim18$\,mag), down to the faint end of the MS. Two fiducial lines (shown in green) were defined to enclose the cluster MS. Stars lying between the two fiducials are shown as black points, while stars outside this region are shown as orange crosses. 

Panel\,(b) displays the corresponding LF. The black histogram shows the raw number counts of the stars selected in panel\,(a) as a function of $m_{\rm F444W}$, using bins of 0.5\,mag. The blue histogram shows the same counts corrected for completeness. The completeness as a function of $m_{\rm F444W}$ is overplotted as a red curve, with its values reported on the upper horizontal axis. The error bars associated with the raw counts are Poisson uncertainties, while those on the completeness-corrected LF represent Poisson errors propagated through the completeness correction. Over this magnitude range, the LF shows a rise that reaches a maximum around $m_{\rm F444W}\sim19.75$\,mag, followed by a decline toward the faint end of the MS. As discussed in Section\,\ref{sec:convective_boundary}, this peak is likely associated with the onset of full convective interiors in low-mass stars. In particular, the change in slope of the mass-luminosity relation associated with this transition produces an accumulation of stars around this magnitude, resulting in the observed peak in the LF.

\begin{centering} 
\begin{figure}
 \includegraphics[width=\columnwidth]{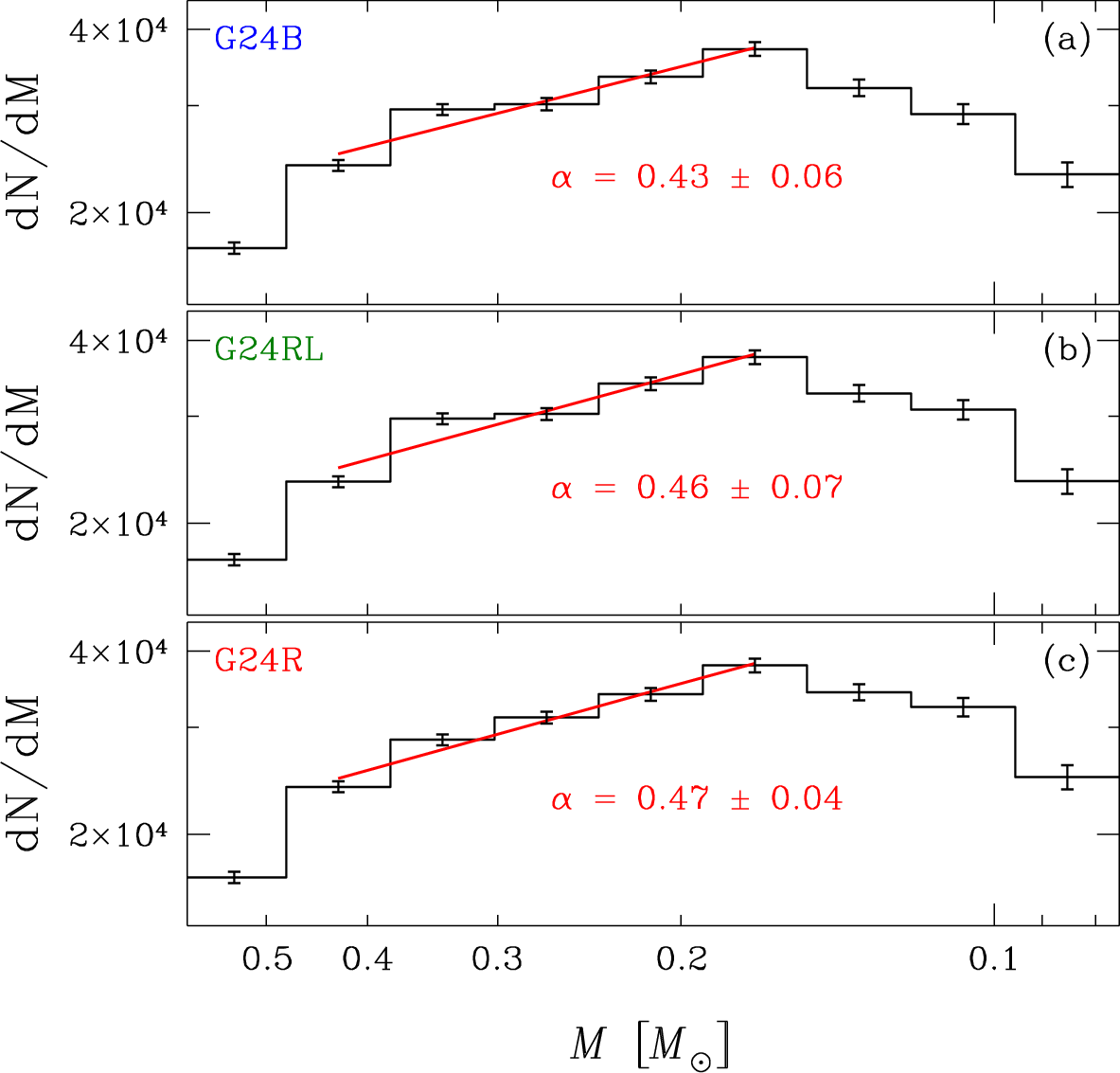}
 \caption{Mass function of the MS and lower MS of 47\,Tuc derived using the three isochrone models: \texttt{G24B} (panel\,a), \texttt{G24RL} (panel\,b), and \texttt{G24R} (panel\,c). Black histograms represent the completeness-corrected number of stars per unit mass, while the red lines indicate the best-fitting power-law relations over the fitted mass range. The corresponding power-law slopes $\alpha$ are reported in each panel.} 
 \label{mf} 
\end{figure} 
\end{centering}

We adopted the isochrone models presented in Fig.\,\ref{iso} to derive the MF for the same sample of stars selected in Fig.\,\ref{lf}. Stellar masses were obtained by converting the $m_{\rm F444W}$ magnitudes using the corresponding mass–luminosity relations. The resulting stellar masses span the range $0.075 \lesssim M/M_\odot \lesssim 0.60$. The sample was then divided into logarithmic mass bins, and the number of stars in each bin was evaluated after correcting for completeness. The resulting MFs are shown in Fig.\,\ref{mf} for the three different sets of isochrones (\texttt{G24B} in panel\,a, \texttt{G24RL} in panel\,b, and \texttt{G24R} in panel\,c). In this figure and in all subsequent LF and MF plots, the error bars represent Poisson uncertainties propagated through the completeness correction. The three MFs are broadly consistent with one another over the explored mass range, indicating that the overall MF shape is only weakly dependent on the adopted mass–luminosity relation.

All three MFs display a similar behaviour: a pronounced gap in the highest-mass bin suggesting a steepening of the MF at higher masses, followed by a smooth increase in the number of stars toward lower masses, a break around $\sim 0.15\,M_{\odot}$, and a subsequent decline in the last three bins at the lowest masses. The break around $\sim0.15\,M_{\odot}$ is qualitatively similar to that observed in a recent JWST-based study of $\omega$\,Centauri \citep{2025A&A...701A.169S}, where the MF was found to deviate from a single power-law form over a comparable mass interval, with a break occurring at a similar mass ($0.17$ -- $0.26\,M_{\odot}$ depending on the adopted model).

To quantify the overall slope of the MF, we fitted the observed binned MFs shown in Fig.\,\ref{mf}, with a single power-law of the form $\xi (M) \propto M^{-\alpha}$, where $\xi(M)$ is the number of stars per unit mass and $\alpha$ is the power-law slope. The fit was performed over the mass range $0.15 \lesssim M/M_\odot \lesssim 0.48$, excluding the highest-mass bin because of its proximity to the onset of saturation, which could introduce systematic uncertainties in the determination of the best-fitting slope, and the three lowest-mass bins, which display an irregular behaviour.

The resulting best-fitting power-law relations are shown as red lines in Fig.\,\ref{mf}, and the corresponding values of $\alpha$ are reported in each panel. The inferred slopes are very similar for the three sets of isochrones and are consistent within the uncertainties, and are in good agreement, within the errors, with the radial variation of the present-day MF inferred for 47\,Tuc \citep[e.g.][]{2018MNRAS.478.1520B,2019MNRAS.482.5138B}, which predicts $\alpha\sim0.7$--1.0 at radial distances comparable to those sampled by our field. The slightly smaller values found here likely reflect the different mass range probed by our analysis.

\begin{centering} 
\begin{figure}
 \includegraphics[width=\columnwidth]{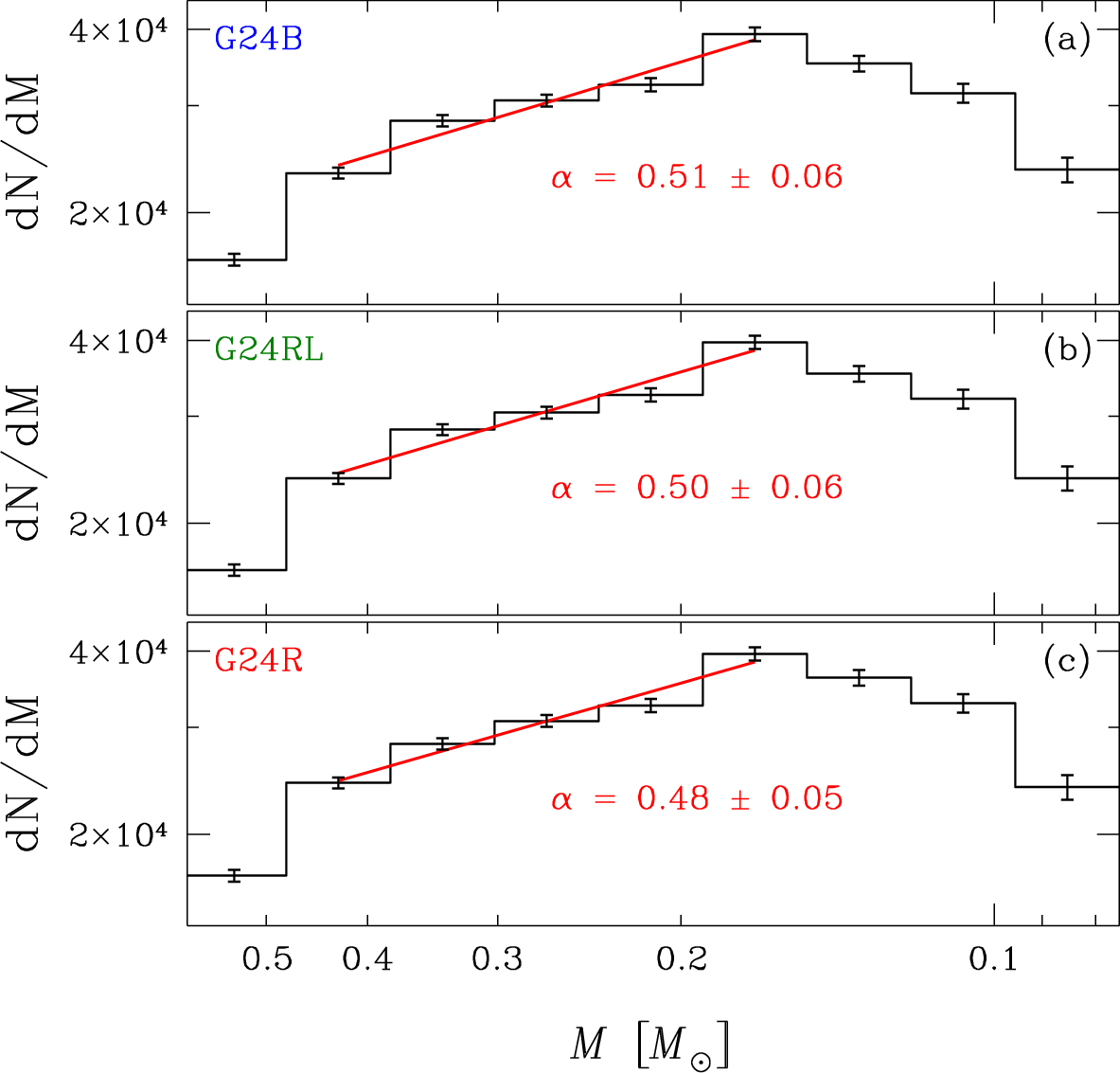}
 \caption{Same as Fig.\,\ref{mf}, but with stellar masses inferred from the $m_{\rm F090W}$–based mass–luminosity relations.} 
 \label{mf090} 
\end{figure} 
\end{centering}

To assess the consistency of the derived MFs with the adopted stellar models and photometric band, we also derived the MFs using stellar masses inferred from the $m_{\rm F090W}$ magnitudes, adopting the corresponding mass–luminosity relations. The resulting MFs are shown in Fig.\,\ref{mf090}. These MFs closely resemble those obtained using $m_{\rm F444W}$, including the presence of a gap at the high-mass end and a change in slope toward lower masses. The best-fitting power-law slopes derived from the $m_{\rm F090W}$–based MFs are similar to, and compatible within the uncertainties with, those obtained from the $m_{\rm F444W}$–based analysis. The derived MF is only weakly sensitive to both the adopted isochrone model and the photometric band used to infer stellar masses.

\begin{centering} 
\begin{figure*}
 \includegraphics[width=\textwidth]{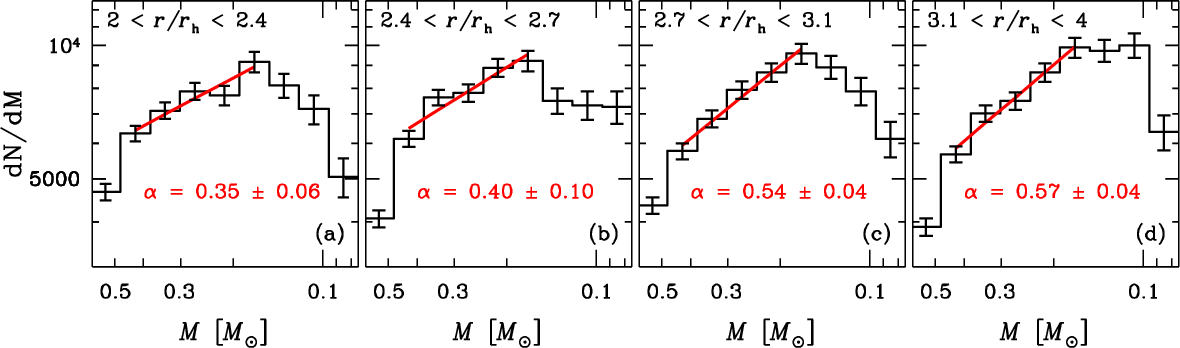}
 \caption{Radial variation of the MF. The MF is shown in four equally populated radial bins, with the corresponding radial intervals indicated in the upper-left corner of each panel. The best-fitting power-law slopes $\alpha$ are reported in the corresponding panels.} 
 \label{mf_bin} 
\end{figure*} 
\end{centering}

To investigate possible radial variations of the MF, we divided the stellar sample into four equally populated radial bins (of $\sim$2,250 stars per bin), spanning the intervals $2 \lesssim r/r_{\rm h} \lesssim 2.4$, $2.4 \lesssim r/r_{\rm h} \lesssim 2.7$, $2.7 \lesssim r/r_{\rm h} \lesssim 3.1$, and $3.1 \lesssim r/r_{\rm h} \lesssim 4$. For each radial bin, we derived the MF following the same procedure adopted for the global MF. Stellar masses were estimated from the $m_{\rm F444W}$ magnitudes using the \texttt{G24R} mass–luminosity relation, as this model provides the best overall match to the observed MS over the explored magnitude range (see Fig.\,\ref{iso}). The resulting radial MFs are shown in Fig.\,\ref{mf_bin}. In all radial bins, the MFs exhibit the same characteristic morphology already observed for the global MF: a pronounced gap at the high-mass end, followed by a smooth increase toward lower masses up to a well-defined maximum, and a subsequent decline in the last three bins at the lowest masses.

As for the global case, the MF in each radial bin was fitted with a single power-law over the same mass range, $0.15 \lesssim M/M_\odot \lesssim 0.48$, excluding the highest-mass bin and the three lowest-mass bins that display irregular behaviour. The best-fitting slopes $\alpha$ are reported in each panel of Fig.\,\ref{mf_bin}. We find that the MF slope increases with increasing distance from the cluster centre, from $\alpha = 0.37 \pm 0.06$ and $\alpha = 0.40 \pm 0.10$ in the two innermost bins, to $\alpha = 0.56 \pm 0.03$ and $\alpha = 0.58 \pm 0.04$ in the outer bins. This behaviour indicates a mild but systematic steepening of the MF with radius and is qualitatively consistent with expectations from mass segregation driven by two-body relaxation, whereby lower-mass stars preferentially migrate towards the cluster's outer regions.

\begin{table*}
\caption{Boundaries of the transition between fully convective stellar interiors and interiors with persistent radiative zones.}
\label{table:convective_gap}
\centering
\begin{tabular}{l r r r r r r r r r r}
\hline
\hline
Isochrone & \multicolumn{5}{c}{Faint edge} & \multicolumn{5}{c}{Bright edge} \\
& $M$ & F090W & F150W2 & F322W2 & F444W & $M$ & F090W & F150W2 & F322W2 & F444W  \\
& [$M_{\odot}$] & [mag] & [mag] & [mag] & [mag] & [$M_{\odot}$] & [mag] & [mag] & [mag] & [mag]  \\
\hline
   \texttt{G24B} & $0.333$ & $21.60$ & $20.51$ & $19.94$ & $19.67$ & $0.337$ & $21.57$ & $20.49$ & $19.91$ & $19.65$ \\
   \texttt{G24RL} & $0.332$ & $21.62$ & $20.50$ & $19.91$ & $19.67$ & $0.336$ & $21.58$ & $20.46$ & $19.87$ & $19.63$\\
   \texttt{G24R} & $0.331$ & $21.59$ & $20.44$ & $19.81$ & $19.63$ & $0.335$ & $21.57$ & $20.42$ & $19.79$ & $19.60$\\
\hline
\end{tabular}
\end{table*}

\subsubsection{A break in the mass function}\label{sec:convective_boundary}

In this section, we assess whether the apparent MF break near $\sim0.15\,M_{\odot}$ is statistically significant and whether it could plausibly arise from a systematic error in our analysis. The break corresponds to a well-defined peak in the MF (see Figs.\,\ref{mf} and \ref{mf090}). Specifically, all bins above the break show a monotonic increase in star counts with stellar mass, while all bins below the break show a monotonic decrease. These trends persist across all three isochrones employed in this study and both photometric bands. Consequently, if the break were driven by a systematic effect, that effect would need to be strongly localised near $\sim0.15\,M_{\odot}$; a smooth, extended bias across a wider mass range would likely produce a more gradual and less consistent transition.

We first examine potential systematic errors arising from the use of mass-luminosity relationships with mismatched chemical composition when converting observed magnitudes into stellar masses. As shown in Fig.\,\ref{iso}, the low MS exhibits a substantial spread in colour, commonly attributed to light-element abundance variations associated with multiple stellar populations. By contrast, the isochrones from \citet{2024ApJ...961..139G} adopted in this work span only a small fraction of the observed colour spread, as they vary solely in $[\mathrm{O/Fe}]$, which is likely insufficient to capture the full structure of the $m_{\rm F090W}-m_{\rm F444W}$ CMD. In our analysis in Section~\ref{obs_lum_MF}, we have implicitly assumed that the mass–luminosity relationship is independent of stellar colour. Equivalently, this corresponds to extrapolating theoretical isomass lines as horizontal lines in the CMD space, so that stellar masses are inferred from magnitude alone, irrespective of colour offsets from the isochrone.

Fig.\,\ref{iso} also shows that the CMD morphology changes significantly along the low MS. Specifically, the sequence transitions from nearly vertical to a much shallower slope, and the onset of this transition occurs at a stellar mass comparable to that of the MF break. It is therefore important to assess whether our method of extrapolating isomass lines across the observed colour spread could introduce a systematic bias in the inferred masses, and in particular whether such bias could vary rapidly with mass, thereby mimicking the apparent MF break.

To explore this possibility, we re-derived stellar masses using a range of isomass extrapolation schemes. For a given stellar mass, we identified the corresponding points on the three adopted isochrones (\texttt{G24B}, \texttt{G24RL}, and \texttt{G24R}) and constructed isomass lines by interpolating and extrapolating between these points in the CMD space. Specifically, we considered three approaches: \textit{(1)} linear extrapolation, obtained by fitting a straight line through the three points; \textit{(2)} quadratic extrapolation, obtained by fitting a parabola; and \textit{(3)} piecewise-linear extrapolation. Stellar masses were then assigned to each star using the isomass-matching method from \citet{2024AN....34540018S}, in which we identify the isomass line that passes closest to the star's CMD location. We excluded all stars with estimated masses over $0.4\,M_{\odot}$, as the isochrones overlap in that region of the CMD, resulting in the polynomial fits being ill-conditioned (i.e. small perturbations in polynomial coefficients have a large effect on estimated masses). Because this mass range lies well above the location of the MF break, this cut does not affect our conclusions.

We then fitted a broken power-law MF to the inferred masses using the histogram-free likelihood maximisation method \citep{2025A&A...701A.169S,Roman_GC_ages}. The unnormalised likelihood, $\mathcal{L}$, is defined as

\begin{equation}
    \log(\mathcal{L})=\sum_i \frac{\log\left(\begin{cases}
        M_i^{-\alpha_\mathrm{h}},& \text{if } M_i>M_\mathrm{bp}\\
        M_i^{-\alpha_\mathrm{l}},& \text{if } M_i\leq M_\mathrm{bp}\\
    \end{cases}\right)}{\mathrm{comp}_i}
    \label{eq:likelihood}
\end{equation}

where $M_i$ are the inferred stellar masses and $\mathrm{comp}_i$ are the corresponding completeness estimates (determined by interpolating the completeness curve in Fig\,\ref{lf} to the F444W magnitude of the star). The free parameters of the model are the power-law break mass, $M_\mathrm{bp}$; and the MF slopes at masses larger and smaller than $M_\mathrm{bp}$: $\alpha_\mathrm{h}$ and $\alpha_\mathrm{l}$, respectively. We also fitted an unbroken power-law MF to each set of inferred masses by fixing $\alpha_\mathrm{h}=\alpha_\mathrm{l}=\alpha$.

\begin{centering} 
\begin{figure}
 \includegraphics[width=\columnwidth]{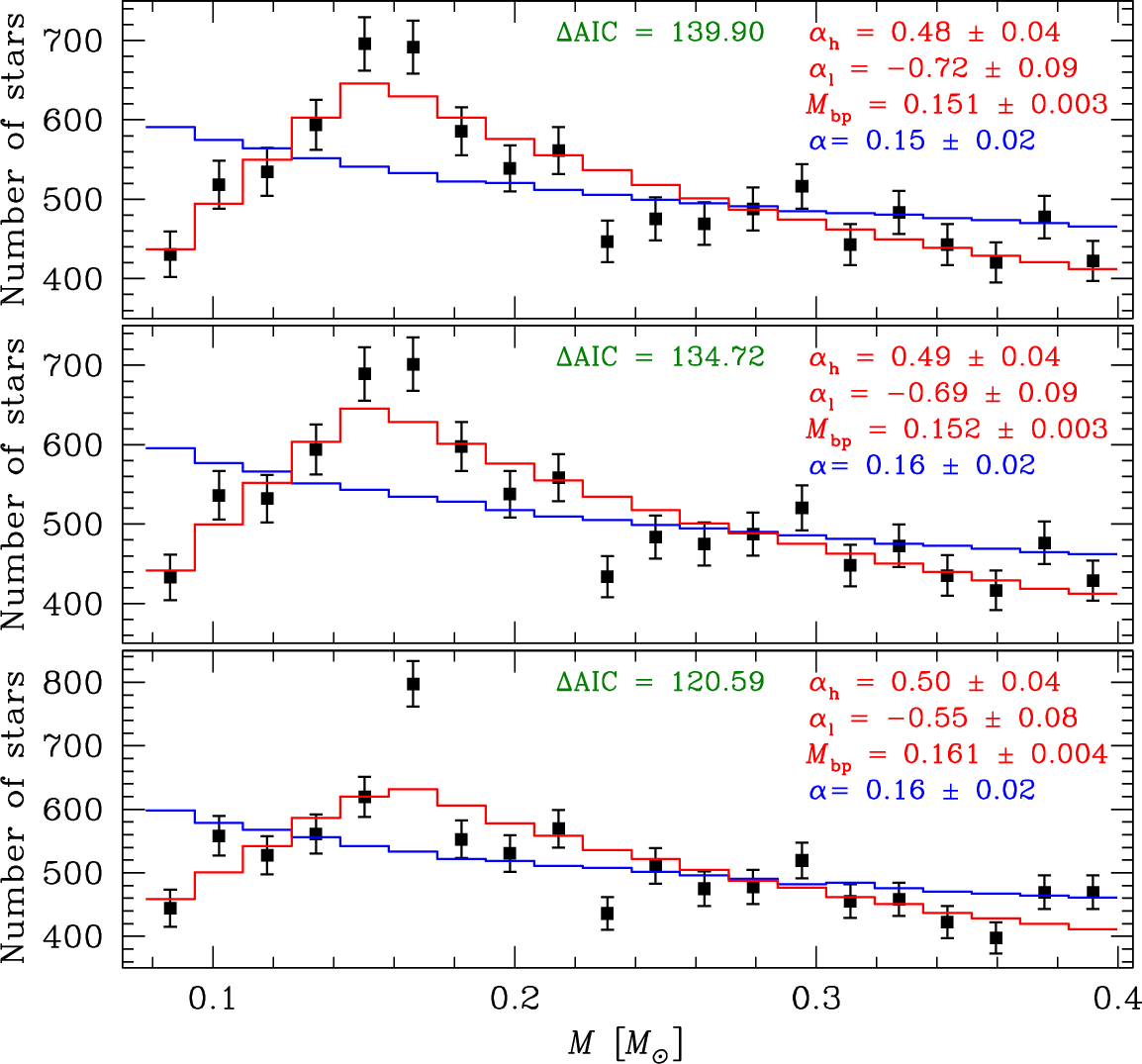}
 \caption{MFs of 47\,Tuc inferred from our observations under different isomass extrapolation schemes (\textit{top:} linear, \textit{middle:} piece-wise linear, \textit{bottom:} quadratic). Star counts have been corrected for completeness. The error bars are completeness-scaled Poisson uncertainties. The best-fit broken (red) and unbroken (blue) power-law models are shown together with their corresponding parameters: the power-law break mass, $M_\mathrm{bp}$, and the MF slopes at masses larger and smaller than $M_\mathrm{bp}$, $\alpha_\mathrm{h}$ and $\alpha_\mathrm{l}$, respectively. The differences between the unbroken and broken MF model Akaike Information Criterion ($\Delta\mathrm{AIC}$) are also shown. Note that while the MFs are presented as histograms here, model fitting was carried out without binning using likelihood maximisation.} 
 \label{fig:MCMC_MF} 
\end{figure} 
\end{centering}

The likelihood was sampled using the Goodman-Weare \citep{MCMC} Markov Chain Monte Carlo (MCMC) algorithm implemented in the \texttt{emcee} Python package \citep{emcee_2013}. We used $32$ MCMC walkers and $5000$ steps per walker. The best-fit MF parameters were extracted from the MCMC step with the highest likelihood in the chain, while the corresponding errors were estimated as the standard deviations of the MCMC posteriors. The broken and unbroken power-law fits for all three isomass interpolation schemes as well as the best-fit parameters are shown in Fig.\,\ref{fig:MCMC_MF}.

For all isomass extrapolation schemes shown in Fig.\,\ref{fig:MCMC_MF}, the best-fit power-law slope above the break ($\alpha_\mathrm{h}$) remains consistent with the range of values derived in Section~\ref{obs_lum_MF}. The figure also reports the difference in the Akaike Information Criterion (AIC; \citealt{AIC}) between the unbroken and broken power-law models. In every case, this difference is large and positive, indicating that the broken power-law model is strongly preferred despite its additional degrees of freedom.

In the case of quadratic extrapolation, however, the AIC difference is driven primarily by a narrow mass interval between $0.15$ and $0.17\ M_\odot$, corresponding to a single bin in the lower panel of Fig.\,\ref{fig:MCMC_MF} that contains a pronounced excess of stars relative to neighbouring bins. When stars in this interval are excluded, the AIC difference decreases by a factor of three, and the best-fit break mass $M_\mathrm{bp}$ shifts to substantially lower values, no longer aligning with the candidate MF break. Under these conditions, our observations are most consistent with an unbroken power law between $0.1$ and $0.4\ M_\odot$.

Importantly, this behaviour is unique to the quadratic extrapolation scheme: for the other extrapolation methods, the preference for a broken power-law remains robust even when analogous narrow mass intervals are removed. This indicates that attributing the MF break to a systematic error in isomass extrapolation would imply the presence of a substantial overdensity of stars confined to a very narrow mass range.

Finally, we consider the possibility that the observed MF break arises from a systematic error in the mass–luminosity relationships themselves. As argued above, reproducing the sharpness of the break would require such an error to be strongly localised near $\sim0.15\,M_{\odot}$. In practise, this would imply a discontinuity or rapid change in the slope of the adopted mass-luminosity relationships at this mass.

A known physical feature of the low MS that may be responsible for such a discontinuity is the transition from full convection to stellar interiors that develop a radiative zone. This transition is also associated with a change in interior opacity due to dissociation of molecular hydrogen \citep{1997MNRAS.287..402K}. In standard models, it occurs at masses  $>0.3\,M_{\odot}$, although it has been predicted to shift to lower masses in GC environments \citep{2021A&A...650A.184M}.

To determine the location of this transition in the models adopted here, we calculated new evolutionary models using \texttt{MESA} (\texttt{M}odules for \texttt{E}xperiments in \texttt{S}tellar \texttt{A}strophysics) code, version \texttt{23.05.1} \citep{MESA,MESA_2,MESA_3,MESA_4,MESA_5}, following the method in \citet{SANDee}. We carried out this analysis separately for each of the isochrones in Fig.\,\ref{iso}, using the pre-tabulated atmospheric boundary conditions from \citet{2024ApJ...961..139G}.

We chose the masses of new \texttt{MESA} models using the bisection root-finding method in order to identify two characteristic stellar masses at the age of $11.5\ \mathrm{Gyr}$. For each of the three chemical compositions corresponding to the G24B, G24RL, and G24R isochrones, we computed a sequence of models around the expected convective boundary and examined their internal structure. The faint edge of the transition region was defined as the highest stellar mass that remains fully convective throughout its evolution, whereas the bright edge was defined as the lowest stellar mass that develops a stable radiative zone and does not subsequently return to a fully convective configuration. These intermittent episodes, known as the convective kissing instability \citep{convective_kissing}, occur in stars within the transition regime between fully convective structures and those with a persistent radiative zone. Stars with masses and magnitudes within this interval undergo the convective kissing instability and alternate between the two configurations during their evolution. Our results are summarised in Table,\ref{table:convective_gap}. The corresponding magnitudes were obtained by interpolating the adopted isochrones at the masses determined from the \texttt{MESA} calculations. For all cases shown in the table, the predicted mass of the convective boundary exceeds our best-fit $M_\mathrm{bp}$ (Fig.,\ref{fig:MCMC_MF}) by over a factor of two. We may therefore safely conclude that the observed MF break is unrelated to the convective discontinuity in the mass-luminosity relationship.

The tests presented in this section indicate that the MF break near $\sim0.15\,M_{\odot}$ is statistically significant and cannot be readily attributed to systematic uncertainties in the modelling. In particular, reproducing the observed feature via systematics would require a substantial ($\sim25\%$) overdensity of stars confined to the narrow mass interval $0.15<M<0.17\,M_{\odot}$, which appears unlikely to be physical. This feature is qualitatively similar to a comparable break reported in the GC $\omega$\,Centauri \citep{2025A&A...701A.169S}. If confirmed, it may point to a presently unrecognised mechanism operating in the star formation or dynamical evolution of GCs.

The presence of the MF break also has direct implications for the expected number of detectable BDs in 47\,Tuc. Extrapolating the best-fit broken and unbroken MF models using the masses derived from the linear isomass extrapolation (upper panel of Fig.\,\ref{fig:MCMC_MF}) into the substellar regime ($M<0.074\,M_\odot$), and restricting to magnitudes brighter than the faintest theoretical completeness limit in Fig.\,\ref{completeness_vs_predicted} ($m_{\rm F444W} \approx 26$\,mag, corresponding to $\sim0.06\,M_{\odot}$), we estimate that approximately $\approx 130$ and $\approx 75$ BDs should be detectable in future JWST observations of our field under the unbroken and broken MF assumptions, respectively.

\subsubsection{On the faint-end luminosity function}\label{sec:faint_end_LF}

The faint end of the lower MS of 47\,Tuc has been explored with JWST observations starting from the work of \citet{2023MNRAS.521L..39N}, who reported the possible detection of the brightest part of the BD sequence. The LF of this regime was later analysed by \citet{2024ApJ...965..189M}. In that study, the LF was reported to show a distinct upturn at its faint end, occurring after the LF reaches a minimum around $m_{\rm F322W2}\sim25.3$\,mag (see their Fig.\,6). This rise was interpreted as a possible signature of the transition from the stellar to the substellar regime and as evidence for the presence of a population of BDs in the cluster. In a subsequent analysis, \citet{2025A&A...694A..68S} revisited the faint end of the CMD of 47\,Tuc using PM membership derived from two JWST epochs. That work showed that, once PM selections are applied, the majority of the sources fainter than $m_{\rm F322W2}\sim25.5$\,mag \citep[the sources responsible for the apparent upturn of the LF reported by][]{2024ApJ...965..189M} disappear. Only two candidate BDs\footnote{Coordinates: ($\alpha,\delta$) = 00:21:49.6171, $-$72:03:30.226 and 00:22:07.2514, $-$72:03:26.978.}, remain after the PM selection, but their cluster membership is uncertain due to large PM uncertainties and the very low completeness at these magnitudes. The rise in the LF reported by \citet{2024ApJ...965..189M} was therefore interpreted as most likely being driven by contamination from background point-like galaxies that populate the faint end of the CMD when PM information is not available. The analysis presented in this work seems to confirm the same result using an independent JWST dataset. As shown in Fig.\,\ref{lf}, the LF derived from our PM-cleaned sample does not display any increase at the faintest magnitudes. As illustrated in \citet{2025A&A...694A..68S} and in this work, the application of PM selections is crucial to remove the large number of faint contaminants that populate the lower part of the CMD (see, for example, Fig.\,\ref{comp}). A reliable membership determination is essential in the extremely faint regime, where contamination can significantly bias the inferred LFs and MFs.

In a recent paper, \citet{2026A&A...706A.313V} investigated the faint end of the MS of 47\,Tuc by developing stellar evolution models and deriving, with these, the expected LF. Their models predict that the increase in the LF associated with the stellar-to-BD transition should occur around $m_{\rm F322W2}\sim27$\,mag (see the left panel of their Fig.\,5), about $\sim1.5$ magnitudes fainter than the rise observed by \citet{2024ApJ...965..189M}. This discrepancy led the authors to suggest that the disagreement may reflect uncertainties in the theoretical models. However, we note that the observational LF adopted for this comparison (see the left panel of their Fig.\,5) does not include any PM membership selections. As discussed above, the lack of PM selections at these magnitudes leads to significant contamination in the faintest bins of the LF. In particular, most of the sources shown in cyan in their Fig.\,4 below $m_{\rm F322W2}\sim25.5$\,mag, which are responsible for the rise in the LF shown in the left panel of their Fig.\,5, were shown to disappear once a PM membership selection is applied \citep[see][]{2025A&A...694A..68S}. The observational LF employed in both \citet{2024ApJ...965..189M} and \citet{2026A&A...706A.313V} is therefore likely affected by contamination due to the fact that PM information was not taken into account, which produces the rise observed at the faint end. Therefore, the discrepancy between the models and the observed LF in \citet{2026A&A...706A.313V} cannot be interpreted as evidence of problems in the stellar models. Reliable PM membership is essential when estimating the LF at the faint end of the MS. Comparing theoretical models with a LF that does not account for reliable membership determination at these magnitudes can lead to misleading conclusions regarding the behaviour of stellar models near the stellar–substellar boundary.

\section{Summary}\label{Section7}
In this paper, we presented the first results of a series of studies based on JWST observations of the GC 47\,Tuc obtained within programme GO-5896. Exploiting the combination of the new GO-5896 data with earlier JWST observations, we derived high-precision PMs over a multi-year baseline and constructed clean samples of cluster members extending to very faint magnitudes along the MS and lower MS.

We investigated the presence of the low-mass MS kink previously identified in 47\,Tuc \citep{2024ApJ...965..189M,2025A&A...694A..68S} in the $m_{\rm F090W}-m_{\rm F444W}$ CMD, finding no significant signature of the feature. The reduced prominence of this feature in this colour combination are in agreement with its proposed physical interpretation as a consequence of changes in atmospheric chemistry in very low-mass stars, in particular, the transition in carbon chemistry toward a methane-dominated equilibrium.

A comparison with theoretical isochrones shows that the lower MS is well reproduced only by the oxygen-poor \texttt{G24R} model, while the other models fail to match the observed sequence in this regime, as previously reported by \citet{2025A&A...694A..68S}. This suggests that the stellar population near the bottom of the MS is dominated by oxygen-poor stars.

Using artificial star tests, we quantified the completeness of our sample and showed that the requirement of PM-based membership significantly limits the detection of the faintest objects. This constraint sets the effective depth of the dataset and prevents current JWST observations from reliably tracing the BD cooling sequence in 47\,Tuc. Our faintest PM-selected members reach masses of $\sim0.077\,M_\odot$, remaining slightly above the estimated HBL of 47\,Tuc ($\sim0.074\,M_\odot$).

We analysed the LF and MF of 47\,Tuc. The LF shows a rise reaching a maximum around $m_{\rm F444W}\simeq19.75$\,mag, likely associated with the onset of fully convective interiors in low-mass stars, followed by a decline toward the faint end of the MS, with no evidence for an upturn at the lowest luminosities that could be attributed to the presence of BDs, in agreement with the results of a recent study \citep{2025A&A...694A..68S}. The MF exhibits a structured behaviour, characterised by a gap at the high-mass end and a break toward lower masses at $\sim$0.15\,$M_{\odot}$, qualitatively similar to that observed in a recent JWST-based study of $\omega$\,Centauri \citep{2025A&A...701A.169S}. The location of this break does not coincide with the onset of full convection in stellar interiors, which occurs at $\sim0.30$–$0.35\,M_\odot$, significantly above the inferred break mass ($\sim$0.15\,$M_{\odot}$). Fitting a single power-law over a restricted mass interval yields MF slopes that are robust against the choice of isochrone model and photometric band used to estimate stellar masses. By dividing the stellar sample into equally populated radial bins, we investigated the radial dependence of the MF. We find a mild but systematic steepening of the MF slope with increasing distance from the cluster centre, consistent with expectations from mass segregation driven by two-body relaxation. 

Using an isomass-based mass determination and a likelihood-based fitting approach, we find that a broken power-law model is generally preferred over an unbroken one, with a statistically significant improvement in the fit across most of the adopted isomass extrapolation schemes. Extrapolations of the MF into the substellar regime suggest that the expected number of detectable BDs in our field ranges between $\sim75$ and $130$, depending on the assumed MF shape.

Reproducing the observed break without invoking a change in slope would require a substantial overdensity of stars confined to a narrow mass range, which appears unlikely to be physical. However, we cannot exclude that this feature arises from uncertainties in the adopted mass–luminosity relation at the lowest masses ($\lesssim0.15\,M_\odot$). If confirmed, this break in the MF may point to a presently unrecognised mechanism affecting the formation or dynamical evolution of low-mass stars in GCs.

This paper establishes a homogeneous and well-characterised photometric and statistical framework for studying the low-mass stellar content of 47\,Tuc with JWST. The results presented here lay the groundwork for subsequent papers in this series, which will focus on the internal kinematics of the cluster and on the properties of its white dwarf population.

\begin{acknowledgements}
This work is based on observations made with the NASA/ESA/CSA James Webb Space Telescope. The data were obtained from the Mikulski Archive for Space Telescopes at the Space Telescope Science Institute, which is operated by the Association of Universities for Research in Astronomy, Inc., under NASA contract NAS 5-03127 for JWST. These observations are associated with program GO-2559, GO-2560, and GO-5896. The data described here may be obtained from \url{https://doi.org/10.17909/56dr-q737}
\end{acknowledgements}

\bibliographystyle{aa}
\bibliography{main.bib}

\begin{appendix}
\onecolumn
\section{Additional material}\label{Appendix:A}

\begin{centering} 
\begin{figure*}[h!]
 \includegraphics[width=\textwidth]{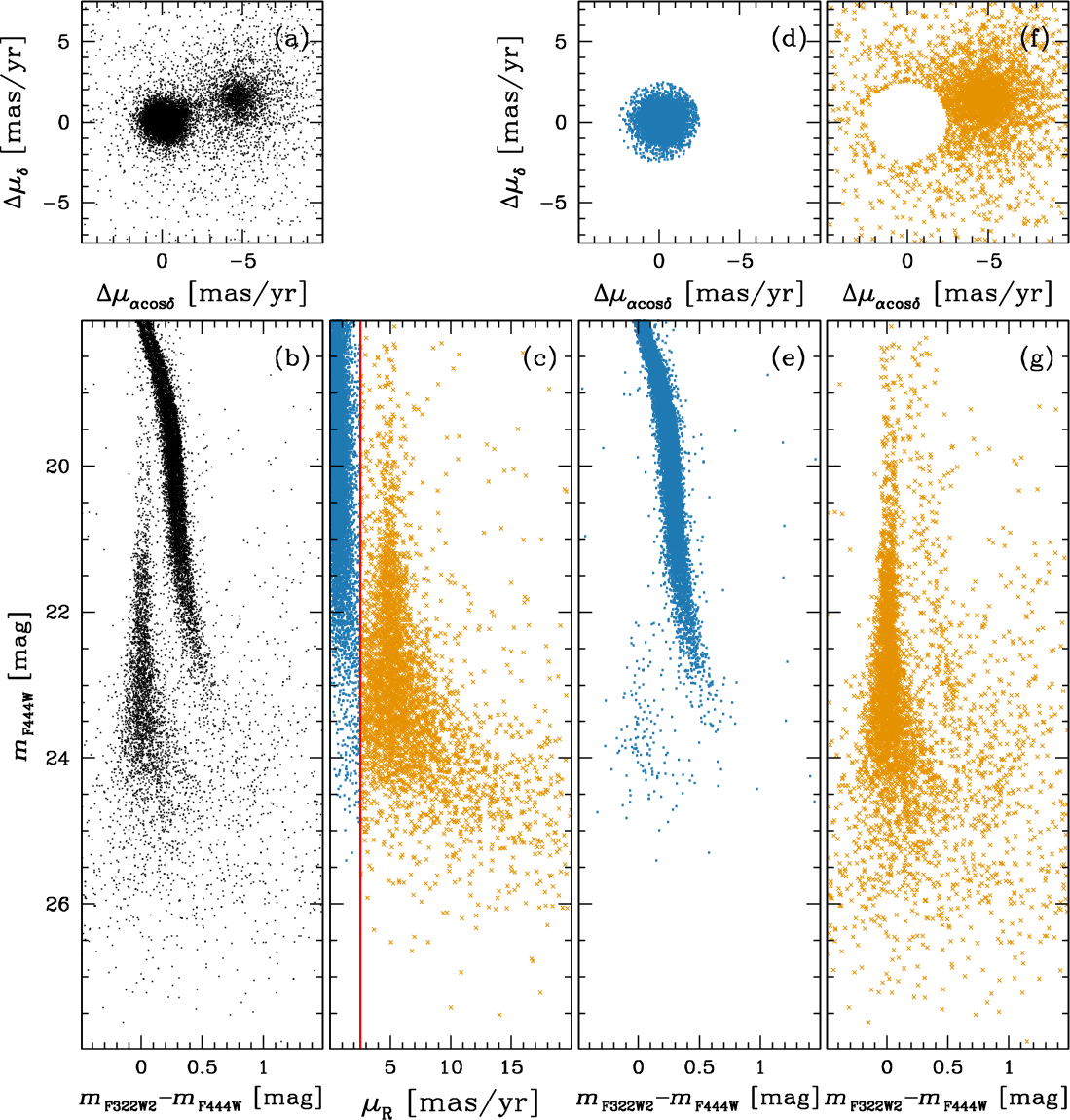}
 \caption{Same as Fig.\,\ref{pm}, but constructed using only the F322W2 and F444W filters.} 
 \label{pm2} 
\end{figure*} 
\end{centering}

\end{appendix}

\end{document}